\documentclass[10pt]{article}
\usepackage[dvips]{color}
\usepackage{epsfig}
\usepackage{amsmath}
\usepackage{graphicx}

\begin{document}
\setcounter{page}{1}

\pagestyle{plain} \vspace{1cm}

\begin{center}
\Large{\bf Dynamics of Generalized Tachyon Field in Teleparallel Gravity}\\
\small \vspace{1cm} {\bf Behnaz Fazlpour $^{a}$
\footnote{b.fazlpour@umz.ac.ir}} and {\bf Ali Banijamali $^{b}$
 \footnote{a.banijamali@nit.ac.ir}} \\
\vspace{0.5cm}   $^{a}$ {\it Department of Physics, Babol Branch,
Islamic Azad University, Babol, Iran\\} \vspace{0.5cm} $^{b}$ {\it
Department of Basic Sciences, Babol University of Technology, Babol,
Iran\\}
\end{center}
\vspace{1.5cm}
\begin{abstract}
We study dynamics of generalized tachyon scalar field  in the
framework of teleparallel gravity. This model is an extension of
tachyonic teleparallel dark energy model which has been proposed in
[26]. In contrast with tachyonic teleparallel dark energy model that
has no scaling attractors, here we find some scaling attractors
which means that the cosmological coincidence problem can be
alleviated. Scaling attractors present for both interacting and
non-interacting dark energy, dark matter cases.\\


\end{abstract}
\newpage
\section{Introduction}
The usual proposal to explain the late-time accelerated expansion of
our universe is an unknown energy component, dubbed as dark energy.
The natural choice and most attractive candidate for dark energy is
the cosmological constant but it is not well accepted because of the
cosmological constant problem [1] as well as the age problem [2].
Thus, many dynamical dark energy models as alternative possibilities
have been proposed. Quintessence, phantom, k-essence, quintom and
tachyon field are the most familiar dark energy models in the
literature (for reviews on dark energy models, see [3]). The tachyon
field arising in the context of string theory [4] and its
application in cosmology both as a source of early inflation and
late-time cosmic acceleration has been extensively studied [5-8].\\
The so-called "Teleparallel Equivalent of General Relativity" or
Teleparallel Gravity was first constructed by Einstein [9-12]. In
this formulation one uses the curvature-less weitzenbock connection
instead of the torsion-less Levi-Civita connection. The relevant
lagrangian in teleparallel gravity is the torsion scalar T which is
constructed by contraction of the torsion tensor. We recall that the
Einstein-Hilbert Lagrangian R is constructed by contraction of the
curvature tensor. Since teleparallel gravity with torsion scalar as
lagrangian density is completely equivalent to a matter-dominated
universe in the framework of general relativity, it can not be
accelerated. Thus one should generalize teleparallel gravity either
by replacing $T$ with an arbitrary function -the so-called $f(T)$
gravity [13-15] or by adding dark energy into teleparallel gravity
allowing also a non-minimal coupling between dark energy and
gravity. Note that both approaches are inspired by the similar
modifications of general relativity i.e. $f(R)$ gravity [16, 17] and
non-minimally coupled dark energy models in the framework of
general relativity [18-20].\\
Recently Geng et al. [21, 22] have included a non-minimal coupling
between quintessence and gravity in the context of teleparallel
gravity. This theory has been called "teleparallel dark energy" and
its dynamics was studied in [23-25]. Tachyonic teleparallel dark
energy is a generalization of teleparallel dark energy by inserting
a non-canonical scalar field instead of quintessence in the action
[26]. Phase-space analysis of this model has been investigated in
[27]. On the other hand, there is no physical argument to exclude
the interaction between dark energy and dark matter. The interaction
between these completely different component of our universe has
same important consequences such as addressing the coincidence
problem [28]. In this paper we consider generalized tachyon field as
responsible for dark energy in the framework of teleparallel
gravity. We will be interested in performing a dynamical analysis of
such a model in FRW space time. In such a study we investigate our
model for both interacting and non-interacting cases. The basic
equations are presented in section 2. In section 3 the evolution
equations are translated in the language of the autonomous dynamical
system by suitable transformation of the basic variables. Subsection
3.1 deals with phase-space analysis as well as the cosmological
implications of the equilibrium points of the model in
non-interacting dark energy dark matter case. In subsection 3.2 an
interaction between dark energy and dark matter has been considered
an critical points and their behavior extracted.
Section 5 is devoted to a short summary of our results.\\

\section{Basic Equations}
Our model is described by the following action as a generalization
of tachyon teleparallel dark energy model [26],

$$S=\int d^{4}xe \Big[\mathcal{L}_{T}+\mathcal{L}_{\varphi}+\mathcal{L}_{m}\Big],$$
$$\mathcal{L}_{T}=\frac{T}{2\kappa^{2}},$$
\begin{equation}
\mathcal{L}_{\varphi}=\xi f(\varphi)T-V(\varphi)(1-2X)^{\beta},
\end{equation}
where $e=det(e_{\,\mu}^{\,i})=\sqrt{-g}$ ($e_{\,\mu}^{\,i}$ are the
orthonormal components of the tetrad) while $\frac{T}{2\kappa^{2}}$
is the Lagrangian of teleparallelism with $T$ as the torsion scalar
(for an introductory review of teleparallelism see [11]).
$\mathcal{L}_{\varphi}$ shows a non-minimal coupling of generalized
tachyon field $\varphi$ with gravity in the framework of
teleparallel gravity and
$X=\frac{1}{2}\partial_{\mu}\varphi\partial^{\mu}\varphi$. The
second part in $\mathcal{L}_{\varphi}$ is the Lagrangian density of
the generalized tachyon field which has been studied in Ref [29].
$f(\varphi)$ is the non-minimal coupling function, $\xi$ is a
dimensionless constant measuring the non-minimal coupling and
$\mathcal{L}_{m}$ is the matter Lagrangian. For $\beta=\frac{1}{2}$
our model reduced to tachyonic teleparallel dark energy discussed in
[26]. Here we consider the case $\beta=2$ for two reasons. The first
is that for arbitrary $\beta$ our equations will be very complicated
and one can not solve them analytically and the second is that for
$\beta=2$ we will obtain interesting physical results as we will see
below.\\
Furthermore, due to complexity of tachyon dynamics Ref [30] has
proposed an approach based on a re-definition of the tachyon field
as follows,
\begin{equation}
\varphi\rightarrow\phi=\int d\varphi\sqrt{V(\varphi)}\Leftrightarrow
\partial\varphi=\frac{\partial\phi}{\sqrt{V(\phi)}}.
\end{equation}
In order to obtain a closed autonomous system and perform the
phase-space analysis of the model we apply (2) in (1) for $\beta=2$
that leads to the following action:
\begin{equation}
S=\int d^{4}xe \Big[\frac{T}{2\kappa^{2}}+\xi
f(\phi)T-V(\phi)\big(1-\frac{2X}{V(\phi)}\big)^{2}
+\mathcal{L}_{m}\Big].
\end{equation}
In a spatially-flat FRW space-time,
\begin{eqnarray}
ds^{2}=dt^{2}-a^{2}(t)(dr^{2}+r^{2}d\Omega^{2}),
\end{eqnarray}
and a vierbein choice of the form $e^{i}_{\mu}=diag(1,a,a,a)$, the
corresponding Friedmann equations are given by,
\begin{equation}
H^{2}=\frac{1}{3}\big(\rho_{\phi}+\rho_{m}\big),
\end{equation}
\begin{equation}
\dot{H}=-\frac{1}{2}\big(\rho_{\phi}+P_{\phi}+\rho_{m}+P_{m}\big),
\end{equation}
where $H=\frac{\dot{a}}{a}$ is the Hubble parameter and a dot stands
for the derivative with respect to the cosmic time $t$. In these
equations, $\rho_{m}$ and $P_{m}$ are the matter energy density and
pressure respectively.\\
The effective energy density and pressure of generalized tachyon
dark energy read,
\begin{equation}
\rho_{\phi}=
V(\phi)+2\dot{\phi}^{2}-3\frac{\dot{\phi}^{4}}{V(\phi)}-6\xi
H^{2}f(\phi),
\end{equation}
and
\begin{equation}
P_{\phi}=-V(\phi)+2\xi\big(3H^{2}+2\dot{H}\big)f(\phi)+10\xi H
f_{,\phi}\dot{\phi}+\dot{\phi}^{2}\big(2-\frac{\dot{\phi}^{2}}{V(\phi)}\big),
\end{equation}
where $f_{,\phi}=\frac{df}{d\phi}$.\\
The equation of motion of the scalar field can be obtained by
variation of the action (3) with respect to $\phi$,
\begin{equation}
\ddot{\phi}+3\mu^{-2}\nu^{2}H\dot{\phi}+\frac{1}{4}\nu^{2}\big(1+\frac{3\dot{\phi}^{4}}{V^{2}(\phi)}\big)V_{,\phi}+6\xi
\nu^{2}H^{2}f_{,\phi}=-\frac{Q}{\dot{\phi}},
\end{equation}
with $Q$ a general interaction coupling term between dark energy and
dark matter, $\mu=\frac{1}{\sqrt{1-\frac{2X}{V}}}$ and
$\nu=\frac{1}{\sqrt{1-\frac{6X}{V}}}$. In (7), (8) and (9) we have
used the useful relation,
\begin{equation}
T=-6H^{2},
\end{equation}
which simply arises from the calculation of torsion scalar for the
FRW metric (4). The scalar field evolution (9) expresses the
continuity equation for the field and matter as follows
\begin{equation}
\dot{\rho}_{\phi}+3H(1+\omega_{\phi})\rho_{\phi}=-Q,
\end{equation}
\begin{equation}
\dot{\rho}_{m}+3H(1+\omega_{m})\rho_{m}=Q,
\end{equation}
where $\omega_{\phi}=\frac{P_{\phi}}{\rho_{\phi}}$ is the equation
of state parameter of dark energy which is attributed to the scalar
field $\phi$. The barotropic index is defined by
$\gamma\equiv1+\omega_{m}$ with $0<\gamma<2$.\\
Although, dynamics of tachyonic teleparallel dark energy has been
studied in [27], no scaling attractors found. Here we are going to
perform a phase-space analysis of generalized tachyonic teleparallel
dark energy and as we will see below some interesting scaling
attractors appear in such theory.\\
\section{Cosmological Dynamics}
In order to perform phase-space and stability analysis of the model,
we introduce the following auxiliary variables:
\begin{equation}
x\equiv\frac{\dot{\phi}}{\sqrt{V}},\,\,\,y
\equiv\frac{\sqrt{V}}{\sqrt{3}H} ,\,\,\,u\equiv\sqrt{f}.
\end{equation}
The auxiliary variables allow us to straightforwardly obtain the
density parameter of dark energy and dark matter
\begin{equation}
\Omega_{\phi}\equiv \frac{\rho_{\phi}}{3H^{2}}=
\mu^{-2}y^{2}\big(1+3x^{2}\big)-2\xi u^{2},
\end{equation}
\begin{equation}
\Omega_{m}\equiv \frac{\rho_{m}}{3H^{2}}=1-\Omega_{\phi},
\end{equation}
while the equation of state of the field reads
$$\omega_{\phi}\equiv\frac{P_{\phi}}{\rho_{\phi}}$$
\begin{equation}
=\frac{-\mu^{-4}y^{2}+2\xi u\big[\frac{5\sqrt{3}}{3}\alpha x
y+u(1-\frac{2}{3}s)\big]}{\mu^{-2}y^{2}\big(1+3x^{2}\big)-2\xi
u^{2}},
\end{equation}
where $\alpha \equiv\frac{f_{,\phi}}{\sqrt{f}}$ and
\begin{equation}
s=-\frac{\dot{H}}{H^{2}}=(2\xi u^{2}+1)^{-1}\Big[5\sqrt{3}\alpha\xi
u x
y+6\mu^{-2}x^{2}y^{2}-\frac{3}{2}\gamma\mu^{-2}y^{2}\big(1+3x^{2}\big)\Big]+\frac{3\gamma}{2}.
\end{equation}
Another quantities with great physical significance namely the total
equation of state parameter and the deceleration parameter are given
by
\begin{equation}
\omega_{tot}\equiv\frac{P_{\phi}+P_{m}}{\rho_{\phi}+\rho_{m}}=
\mu^{-2} y^{2}\big(4x^{2}-\gamma(1+3x^{2})\big)+2\xi
u\big[\frac{5\sqrt{3}}{3}\alpha x
y+u(\gamma-\frac{2}{3}s)\big]+\gamma-1,
\end{equation}
and
$$q\equiv -1-\frac{\dot{H}}{H^{2}}=\frac{1}{2}+\frac{3}{2}\omega_{tot}$$
\begin{equation}
=\frac{3}{2}\mu^{-2} y^{2}\big(4x^{2}-\gamma(1+3x^{2})\big)+\xi
u\big[5\sqrt{3}\alpha x y+u(3\gamma-2s)\big]+\frac{3\gamma}{2}-1.
\end{equation}
Using auxiliary variables (13) the evolution equations (5), (6) and
(9) can be recast as a dynamical system of ordinary differential
equations
\begin{equation}
x'=\frac{\sqrt{3}}{2}\big[\lambda
x^{2}y+\frac{1}{2}\lambda\nu^{2}(1+3x^{4})y-4\alpha \xi  \nu^{2}u
y^{-1}-2\sqrt{3} \mu^{-2}\nu^{2}x\big]-\hat{Q},
\end{equation}

\begin{equation}
y'=\Big(-\frac{\sqrt{3}}{2}\lambda x y+s\Big)y,
\end{equation}

\begin{equation}
u'=\frac{\sqrt{3}\alpha x y}{2},
\end{equation}
where $\hat{Q}=\frac{Q}{\dot{\phi}H\sqrt{V(\phi)}}$,
$\lambda\equiv-\frac{V_{,\phi}}{\kappa V}$ and prime in equations
(20)-(22) denotes differentiation with respect to the so-called
e-folding time $N=\ln a$.\\
From now we concentrate on exponential scalar field potential of the
form $V=V_{0} e^{-k\lambda\phi}$ and the non-minimal coupling
function of the form $f(\phi)\propto \phi^{2}$. These choices lead
to constant $\lambda$ and $\alpha$ respectively.\\ The next step is
the introduction of interaction term $Q$ to obtain an autonomous
system out of equations (20)-(22). The fixed points
$(x_{c},y_{c},u_{c})$ for which $x'=y'=u'=0$ depend on the choice of
the interaction term $Q$ and two general possibilities will be
treated in the sequel. The stability of the system at a fixed point
can be obtained from the analysis of the determinant and trace of
the perturbation matrix $M$. Such a matrix can be constructed by
substituting linear perturbations $x\rightarrow x_{c}+\delta x,
y\rightarrow y_{c}+\delta y$ and $u\rightarrow u_{c}+\delta u$ about
the critical point $(x_{c}, y_{c}, u_{c})$ into the autonomous
system (20)-(22). The $3 \times3$ matrix $M$ of the linearized
perturbation equations of the autonomous system is shown in appendix
A. Therefor, for each critical point we examine the sign of the real
part of the eigenvalues of $M$. According to the usual dynamical
system analysis, if the eigenvalues are real and have opposite
signs, the corresponding critical point is a saddle point. A fixed
point is unstable if the eigenvalues are positive and it is stable
for
negative real part of the eigenvalues.\\
In the following subsections we will study the dynamics of
generalized tachyon field with different interaction term $Q$.
Without lose of
generality we assume $\gamma=1$ for simplicity.\\

\subsection{The case for $Q=0$}
The first case $Q=0$ clearly means there is no interaction between
dark energy and background matter. In this case, there are two
critical points presented in Table 1. From equations (14) and (16)
one can obtain the corresponding values of density parameter
$\Omega_{\phi}$ and equation of state of dark energy $\omega_{\phi}$
at each point. Also, using equation (19) we can find the condition
required for acceleration $(q<0)$ at each point. These parameters
and conditions have been shown in Table 1. The stability and
existence conditions
of critical points $A_{10}$ and $A_{2}$ are presented in Table 2. We mention that
the corresponding eigenvalues of perturbation matrix $M$ at critical points $A_{10}$ and $A_{2}$
 are considerably involved and here we do not present their explicit expressions but we can find sign of them numerically.\\
\textit{Critical point $A_{1}$}: This critical point is a scaling
attractor if $\xi>0$, $\alpha<0$ and $\lambda<0$. Thus, it can give
the hope alleviate the cosmological coincidence problem. $A_{1}$ is
a saddle point for $\alpha>0$ and $\frac{\xi}{\lambda}>0$.\\
\textit{Critical point $A_{2}$}: $A_{2}$ can also be a scaling
attractor of the model or a saddle point under the same conditions
as for $A_{1}$.\\
In figure 1 we have chosen the values of the parameters $\xi$,
$\lambda$ and $\alpha$, such that $A_{1}$ become a stable attractor
of the model. Plots in figure 1 show the phase-space trajectories on
$x-y$, $x-u$ and $u-y$ planes from left to right respectively. The
same plots are shown in figure 2 for critical point $A_{2}$. Note
that the values of the parameter have chosen in the way that $A_{2}$
become a stable point of the model. In figure 3, the corresponding 3
dimensional phase-space trajectories of the model have been
presented. One can see that $A_{1}$ and $A_{2}$ are stable attractor
of the model in the left and right plots respectively.\\

\begin{table}
\begin{center}

\begin{tabular}{ccccc}
  \hline
   \hline label& $(x_{c}, y_{c}, u_{c})$&$\Omega_{\phi}$& $\omega_{\phi}$  & acceleration \\
   \hline \\ $A_{1}$&$0, 2\sqrt{\lambda_{1}}, \frac{\lambda\lambda_{1}}{2\alpha\xi}$&$4\lambda_{1}\big(1-
   \frac{\lambda^{2}}{8\xi\alpha^{2}}\lambda_{1}\big)$&$\frac{16\xi^{2}\alpha^{4}}{\big(\lambda^{2}\lambda_{1}^{2}+2\xi\alpha^{2}
   \big)(\lambda^{2}\lambda_{1}-8\xi\alpha^{2})}$&
   $\begin{array}{c}
   \lambda^{2}>\frac{\xi\alpha^{2}(4\lambda_{1}-1)}{\lambda_{1}^{2}}\big(1+\sqrt{1-\frac{32\lambda_{1}}{(1-4\lambda_{1})^{2}}}\big)\\
   or\\
\lambda^{2}<\frac{\xi\alpha^{2}(4\lambda_{1}-1)}{\lambda_{1}^{2}}\big(1-\sqrt{1-\frac{32\lambda_{1}}{(1-4\lambda_{1})^{2}}}\big)\\
   \end{array}$\\

   \hline \\ $A_{2}$&$0, 2\sqrt{\lambda_{2}}, \frac{\lambda\lambda_{2}}{2\alpha\xi}$
   &$4\lambda_{2}\big(1-
   \frac{\lambda^{2}}{8\xi\alpha^{2}}\lambda_{2}\big)$&$\frac{16\xi^{2}\alpha^{4}}{\big(\lambda^{2}\lambda_{2}^{2}+2\xi\alpha^{2}
   \big)(\lambda^{2}\lambda_{2}-8\xi\alpha^{2})}$&
   $\begin{array}{c}
   \lambda^{2}>\frac{\xi\alpha^{2}(4\lambda_{2}-1)}{\lambda_{2}^{2}}\big(1+\sqrt{1-\frac{32\lambda_{2}}{(1-4\lambda_{2})^{2}}}\big)\\
   or\\
\lambda^{2}<\frac{\xi\alpha^{2}(4\lambda_{2}-1)}{\lambda_{2}^{2}}\big(1-\sqrt{1-\frac{32\lambda_{2}}{(1-4\lambda_{2})^{2}}}\big)\\
   \end{array}$\\

   \hline
  \hline
\end{tabular}
\caption{Location of the critical points and the corresponding
values of the dark energy density parameter $\Omega_{\phi}$ and
equation of state $\omega_{\phi}$ and the condition required for an
accelerating universe for $Q=0$. Here
$\lambda_{1}=\frac{\alpha\big(4\alpha\xi+\sqrt{16\alpha^{2}\xi^{2}-2\lambda^{2}\xi}\big)}{\lambda^{2}}$
and
$\lambda_{2}=\frac{\alpha\big(4\alpha\xi-\sqrt{16\alpha^{2}\xi^{2}-2\lambda^{2}\xi}\big)}{\lambda^{2}}$.}
\vspace{0.5 cm}
\end{center}
\end{table}

\begin{table}
\begin{center}

\begin{tabular}{ccc}
  \hline
   \hline label  & stability & existence\\
  \hline$A_{1}$&$\begin{array}{c}
                      saddle\,\, point\\
                      if\,\, \alpha>0\,\,\,and\,\,\,\frac{\xi}{\lambda}>0 \\
                      stable\,\, point\\
                         if\,\,\xi>0,\,\,\,
                      \alpha<0\,\,\,and\,\,\,\lambda<0\\
                       \end{array}$&$\begin{array}{c}
                      for\,\,all\,\,\xi<0\\
                      or \\
                      \xi\geq \frac{\lambda^{2}}
                         {8\alpha ^{2}}\\
                            \end{array}$\\

 \hline $A_{2}$&$\begin{array}{c}
                      saddle\,\, point\\
                      if\,\, \alpha<0\,\,\,and\,\,\,\frac{\xi}{\lambda}<0 \\
                      stable\,\, point\\
                         if\,\,\xi>0,\,\,\,
                      \alpha>0\,\,\,and\,\,\,\lambda>0\\
                       \end{array}$&$\begin{array}{c}
                      for\,\,all\,\,\xi<0\\
                      or \\
                      \xi\geq \frac{\lambda^{2}}
                         {8\alpha ^{2}}\\
                            \end{array}$\\

 \hline
 \hline
\end{tabular}
\caption{Stability and existence conditions of the critical points
of the model for $Q=0$.} \vspace{0.5 cm}
\end{center}
\end{table}

\begin{table}
\begin{center}
\begin{tabular}{ccccc}
  \hline
   \hline label& $(x_{c}, y_{c}, u_{c})$&$\Omega_{\phi}$& $\omega_{\phi}$ & acceleration \\
   \hline \\ $B_{1}$&$0, \frac{2\sqrt{2\lambda\alpha\xi\theta_{1}}}{\lambda}, \theta_{1}$&$2\xi\theta_{1}\big(\frac{4\alpha
   }{\lambda}-\theta_{1})$&$\frac{2\big(\xi\theta_{1}^{3}+\frac{2\alpha}{\lambda}\big)}{\big(\theta_{1}-\frac{4\alpha}{\lambda}\big)
   (1+2\xi\theta_{1}^{2})}$
   &$\lambda>\frac{8\alpha(\xi\theta_{1}^{2}-1)}{\theta_{1}(8\xi\theta_{1}^{2}+1)}$
   \\
   \hline \\ $B_{2}$&$0, \frac{2\sqrt{2\lambda\alpha\xi\theta_{2}}}{\lambda}, \theta_{2}$&$2\xi\theta_{2}\big(\frac{4\alpha
   }{\lambda}-\theta_{2})$&$\frac{2\big(\xi\theta_{2}^{3}+\frac{2\alpha}{\lambda}\big)}{\big(\theta_{2}-\frac{4\alpha}{\lambda}\big)
   (1+2\xi\theta_{2}^{2})}$&
   $\lambda>\frac{8\alpha(\xi\theta_{2}^{2}-1)}{\theta_{2}(8\xi\theta_{2}^{2}+1)}$\\
\hline \\ $B_{3}$&$0,
-\frac{2\sqrt{2\lambda\alpha\xi\theta_{1}}}{\lambda},
\theta_{1}$&$2\xi\theta_{1}\big(\frac{4\alpha
   }{\lambda}-\theta_{1})$&$\frac{2\big(\xi\theta_{1}^{3}+\frac{2\alpha}{\lambda}\big)}{\big(\theta_{1}-\frac{4\alpha}{\lambda}\big)
   (1+2\xi\theta_{1}^{2})}$&
   $\lambda>\frac{8\alpha(\xi\theta_{1}^{2}-1)}{\theta_{1}(8\xi\theta_{1}^{2}+1)}$\\
\hline \\ $B_{4}$&$0,
-\frac{2\sqrt{2\lambda\alpha\xi\theta_{2}}}{\lambda},
\theta_{2}$&$2\xi\theta_{2}\big(\frac{4\alpha
   }{\lambda}-\theta_{2})$&$\frac{2\big(\xi\theta_{2}^{3}+\frac{2\alpha}{\lambda}\big)}{\big(\theta_{2}-\frac{4\alpha}{\lambda}\big)
   (1+2\xi\theta_{2}^{2})}$&
   $\lambda>\frac{8\alpha(\xi\theta_{2}^{2}-1)}{\theta_{2}(8\xi\theta_{2}^{2}+1)}$\\

   \hline
  \hline
\end{tabular}
\caption{Location of the critical points and the corresponding
values of the dark energy density parameter $\Omega_{\phi}$ and
equation of state $\omega_{\phi}$ and the condition required for an
accelerating universe for $Q=\beta\kappa\rho_{m}\dot{\phi}$. Here
$\theta_{1}=\frac{\big(4\alpha\xi+\sqrt{16\alpha^{2}\xi^{2}-2\lambda^{2}\xi}\big)}{2\lambda\xi}$
and
$\theta_{2}=\frac{\big(4\alpha\xi-\sqrt{16\alpha^{2}\xi^{2}-2\lambda^{2}\xi}\big)}{2\lambda\xi}$.}
\vspace{0.5 cm}
\end{center}
\end{table}
\newpage
\subsection{The case for $Q=\beta\kappa\rho_{m}\dot{\phi}$}
This deals with the most familiar interaction term extensively
considered in the literature (see e.g. [25, 31-34]). Here $\hat{Q}$
in terms of auxiliary variables is $\hat{Q}=\sqrt{3}\beta
y^{-1}\Omega_{m}$. Inserting such an interaction term in equations
(20)-(22) and setting the left hand sides of the equations to zero
lead to the critical points $B_{1}$, $B_{2}$, $B_{3}$ and $B_{4}$
presented in Table 3. In the same table we have provided the
corresponding values of $\Omega_{\phi}$ and $\omega_{\phi}$ as well
as the condition needed for accelerating universe at each fixed
points.\\
The stability and existence conditions for each point presented in
Table 4. Since the corresponding eigenvalues of the fixed points are complicated we do not give them here but one can
obtain their signs numerically and so concludes about the stability properties of the critical points. \\
\textit{Critical point $B_{1}$}: This point exists for $\lambda>0$
and $\xi\geq\frac{\lambda^{2}}{8\alpha^{2}}$. However it is an
unstable saddle point.\\
\textit{Critical point $B_{2}$}: The critical point $B_{2}$ exists
for $\lambda>0$ and $\xi<0$ or
$\xi\geq\frac{\lambda^{2}}{8\alpha^{2}}$. This point is a scaling
attractor of the model if $\alpha>0$ and $\xi>0$. Figure 4 shows
clearly such a behavior of the model for suitable choices of $\xi$,
$\lambda$ and $\alpha$.\\
\textit{Critical point $B_{3}$}: This point exists for negative
values of $\lambda$ and $\xi$ or when
$\xi\geq\frac{\lambda^{2}}{8\alpha^{2}}$. Also, it is a stable point
if $\alpha<0$ and $\xi>0$ and a saddle point if $\alpha>0$ and
$\xi<0$. The values of parameters have been chosen in figure 5 such
that $B_{3}$ become a attractor of the model as it is clear from
phase-space trajectories.\\
\textit{Critical point $B_{4}$}: The point $B_{4}$ exists for
$\lambda<0$ and $\xi\geq\frac{\lambda^{2}}{8\alpha^{2}}$. It is a
stable point if $\alpha<0$ and $\xi>0$. In figure 6 values of the
parameter $\xi$ and $\alpha$ are those satisfy these constraints and
so $B_{4}$ becomes a attractor point for phase-plane trajectories.
The corresponding 3-dimensional phase-space trajectories of the
model for attractor points $B_{2}$ (left), $B_{3}$ (middle) and
$B_{4}$ (right) are plotted in figure 7.\\

\begin{table}
\begin{center}
\begin{tabular}{ccc}
  \hline
   \hline label  & stability & existence\\
  \hline$B_{1}$&$\begin{array}{c}
                      saddle\,\, point\\
                      if\,\, \alpha>0\,\,\,and\,\,\,\xi>0 \\
                                             \end{array}$&$\lambda>0\,\,\,and\,\,\,\xi\geq \frac{\lambda^{2}}
                         {8\alpha ^{2}}$\\
 \hline $B_{2}$&$\begin{array}{c}
                      saddle\,\, point\\
                      if\,\, \alpha<0\,\,\,and\,\,\,\xi<0 \\
                      stable\,\, point\\
                         if\,\,\alpha>0\,\,\,and\,\,\,
                      \xi>0\\
                       \end{array}$&$\begin{array}{c}
                       \lambda>0\,\,\,and\\
                      for\,\,all\,\,\xi<0\\
                      or \\
                      \xi\geq \frac{\lambda^{2}}
                         {8\alpha ^{2}}\\
                            \end{array}$\\
\hline$B_{3}$&$\begin{array}{c}
                      saddle\,\, point\\
                      if\,\, \alpha>0\,\,\,and\,\,\,\xi<0 \\
                      stable\,\, point\\
                         if\,\, \alpha<0\,\,\,and\,\,\,\xi>0 \\
                       \end{array}$&$\begin{array}{c}
                       \lambda<0\,\,\,and\\
                      for\,\,all\,\,\xi<0\\
                      or \\
                      \xi\geq \frac{\lambda^{2}}
                         {8\alpha ^{2}}\\
                            \end{array}$\\
 \hline $B_{4}$&$\begin{array}{c}
                                            stable\,\, point\\
                         if\,\, \alpha<0\,\,\,and\,\,\,\xi>0 \\
                       \end{array}$&$\lambda<0\,\,\,and\,\,\,\xi\geq \frac{\lambda^{2}}
                         {8\alpha ^{2}}$\\
 \hline
 \hline
\end{tabular}
\caption{Stability and existence conditions of the critical points
of the model for $Q=\beta\kappa\rho_{m}\dot{\phi}$.} \vspace{0.5 cm}
\end{center}
\end{table}

\section{Conclusion}
A model of dark energy with non-minimal coupling of quintessence
scalar field with gravity in the framework of teleparallel gravity
was called teleparallel dark energy [21]. If one replaces
quintessence by tachyon field in such a model then tachyonic
teleparallel dark energy will be constructed [26].\\
Moreover, although dark energy and dark matter scale differently
with the expansion of our universe, according to the observations
[35] we are living in an epoch in which dark energy and dark matter
densities are comparable and this is the well-known cosmological
coincidence problem [24]. This problem can be alleviated in most
dark energy models via the method of scaling solutions in which the
density parameters of dark energy and dark matter are both
non-vanishing over there.\\
In this paper we investigated the phase-space analysis of
generalized tachyon cosmology in the framework of teleparallel
gravity. Our model described by action (1) which generalizes
tachyonic teleparallel dark energy model proposed in [26]. We found
some scaling attractors in our model for the case $\beta=2$. These
scaling attractors are $A_{1}$ and $A_{2}$ when there is no
interaction between dark energy and dark matter. $B_{2}$, $B_{3}$
and $B_{4}$ are scaling attractors in the case that dark energy
interacts with dark matter through the interacting term
$Q=\beta\kappa\rho_{m}\dot{\phi}$. Our results show that generalized
tachyon field represents interesting cosmological behavior in
compare with ordinary tachyon fields in the framework of
teleparallel gravity because there is no scaling attractor in the
latter model. So, generalized tachyon field gives us the hope that
cosmological coincidence problem can be alleviated without
fine-tunings. One can study our model for different kinds of
potential and other famous interaction term between dark energy and dark matter.\\

\newpage
\begin{figure}[htp]
\begin{center}
\includegraphics{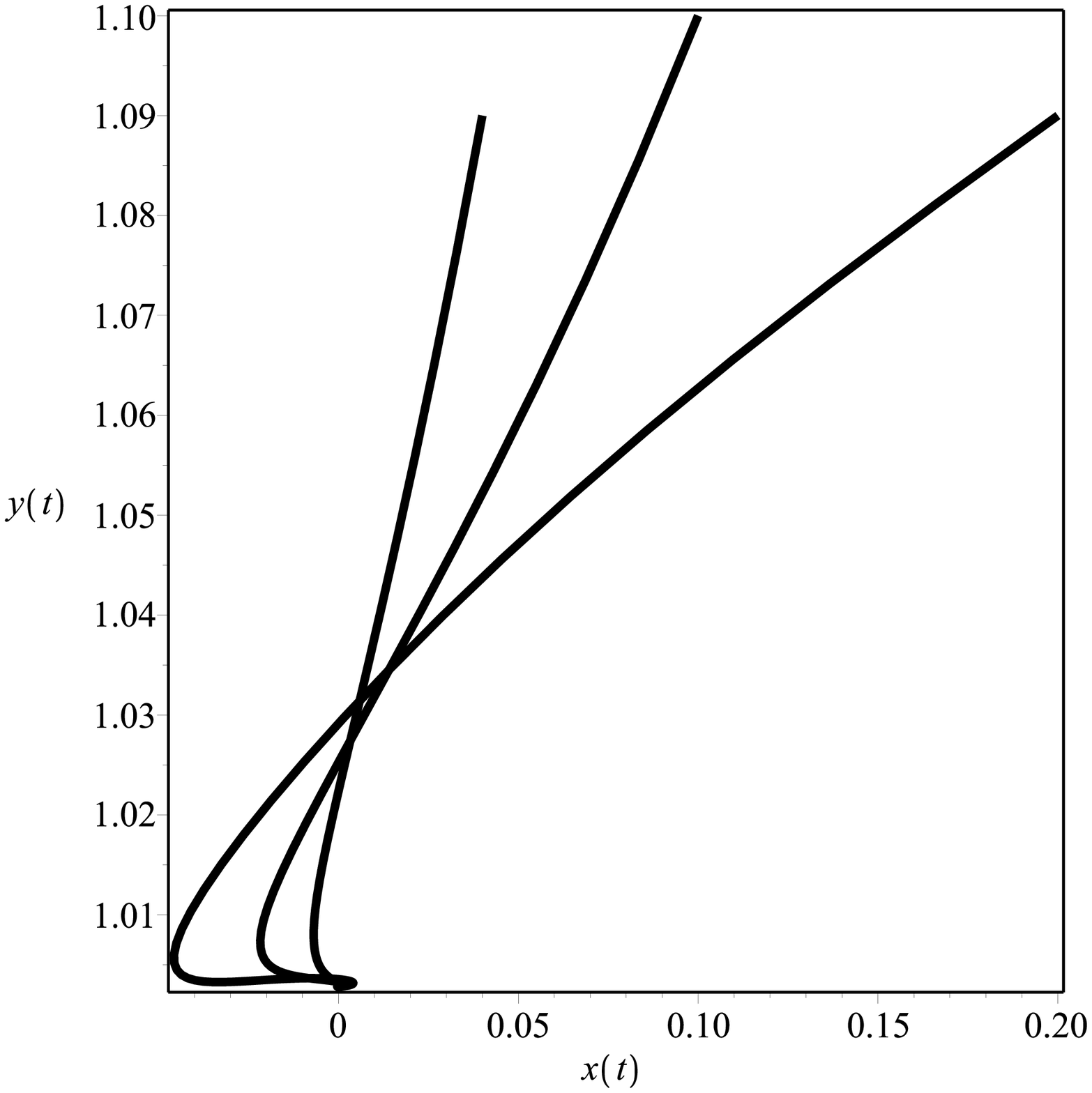} \vspace{2.5cm}\includegraphics{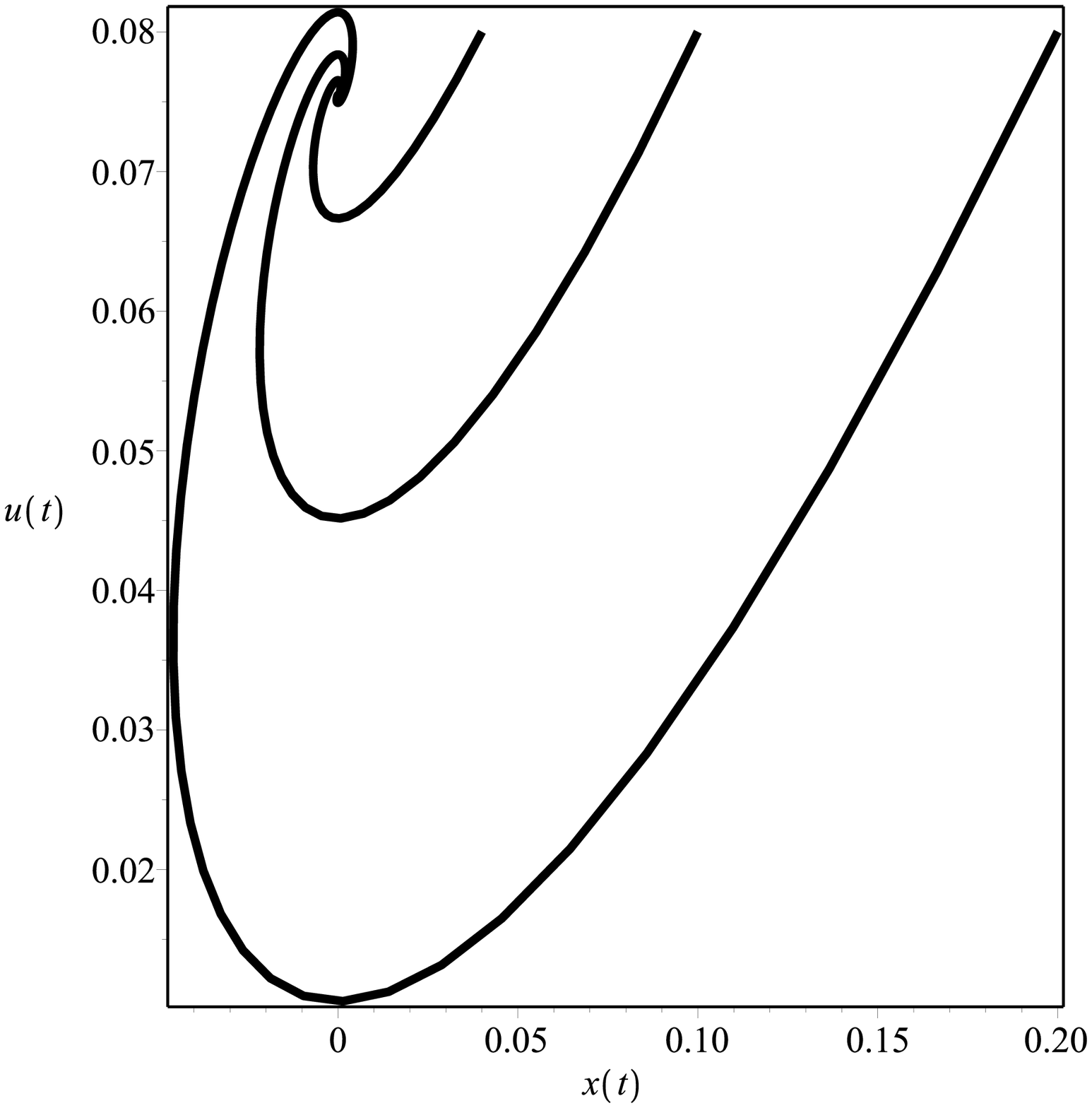}\vspace{2.5cm}\includegraphics{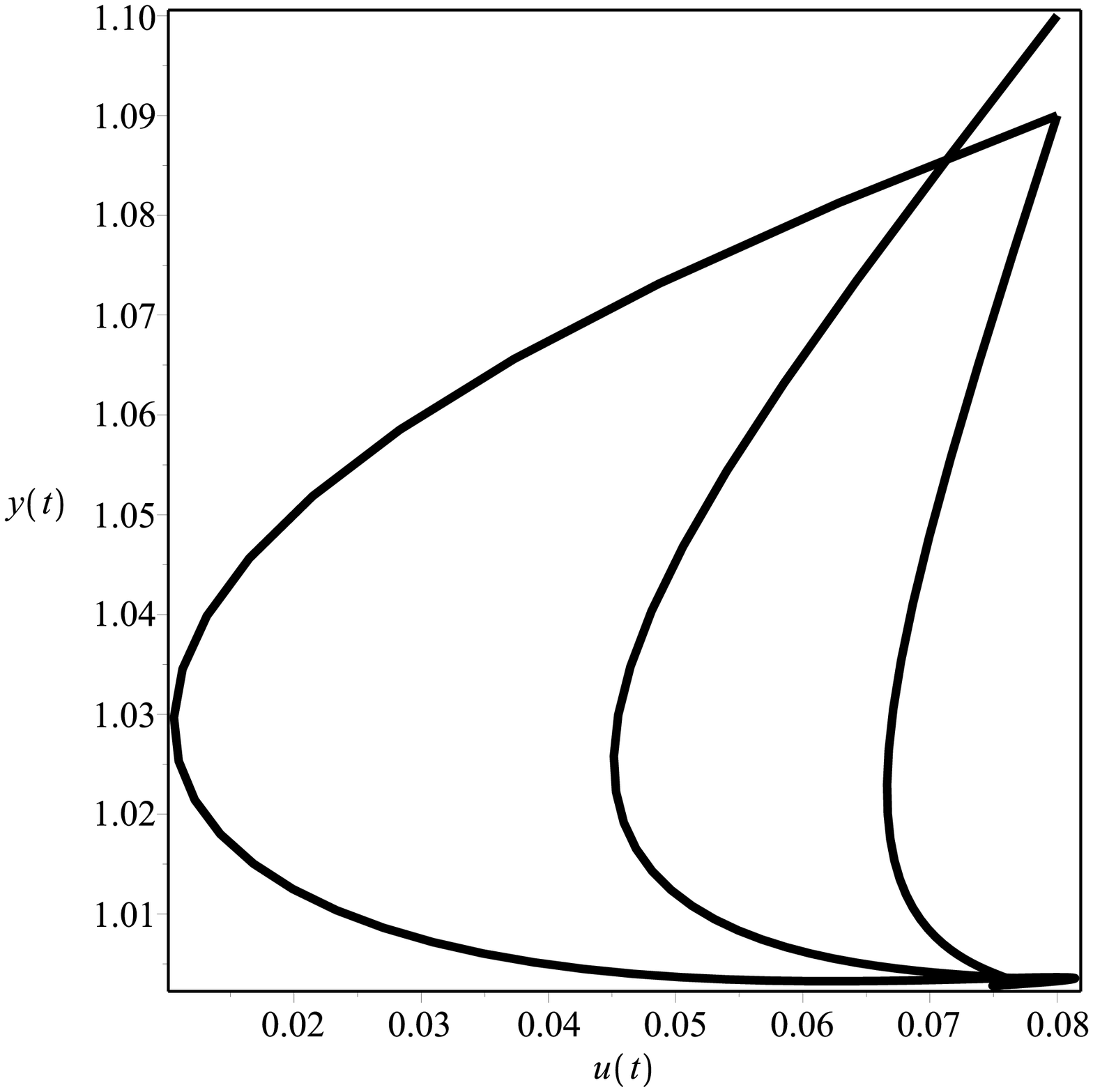} \caption{\small {From
left to right, the projections of the phase-space trajectories on
the $x-y$, $x-u$ and $u-y$ planes with $\xi=0.5$, $\lambda=-0.6$ and
$\alpha=-2$ for $Q=0$. For these values of the parameters, point
$A_{1}$ is a stable attractor of the model. }}

\end{center}
\end{figure}
\vspace{3cm}
\begin{figure}[htp]
\begin{center}
\includegraphics{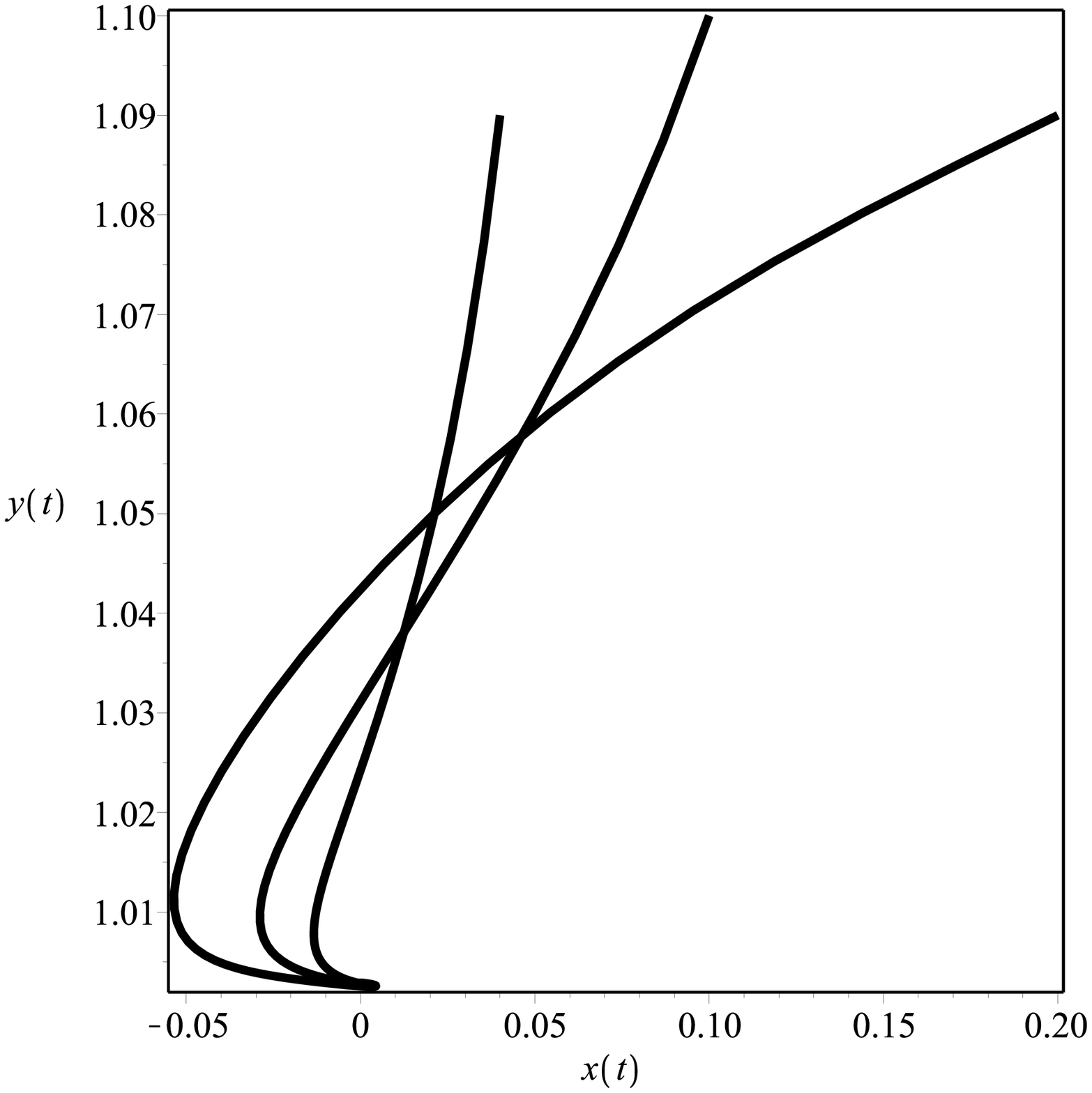} \vspace{2.5cm}\includegraphics{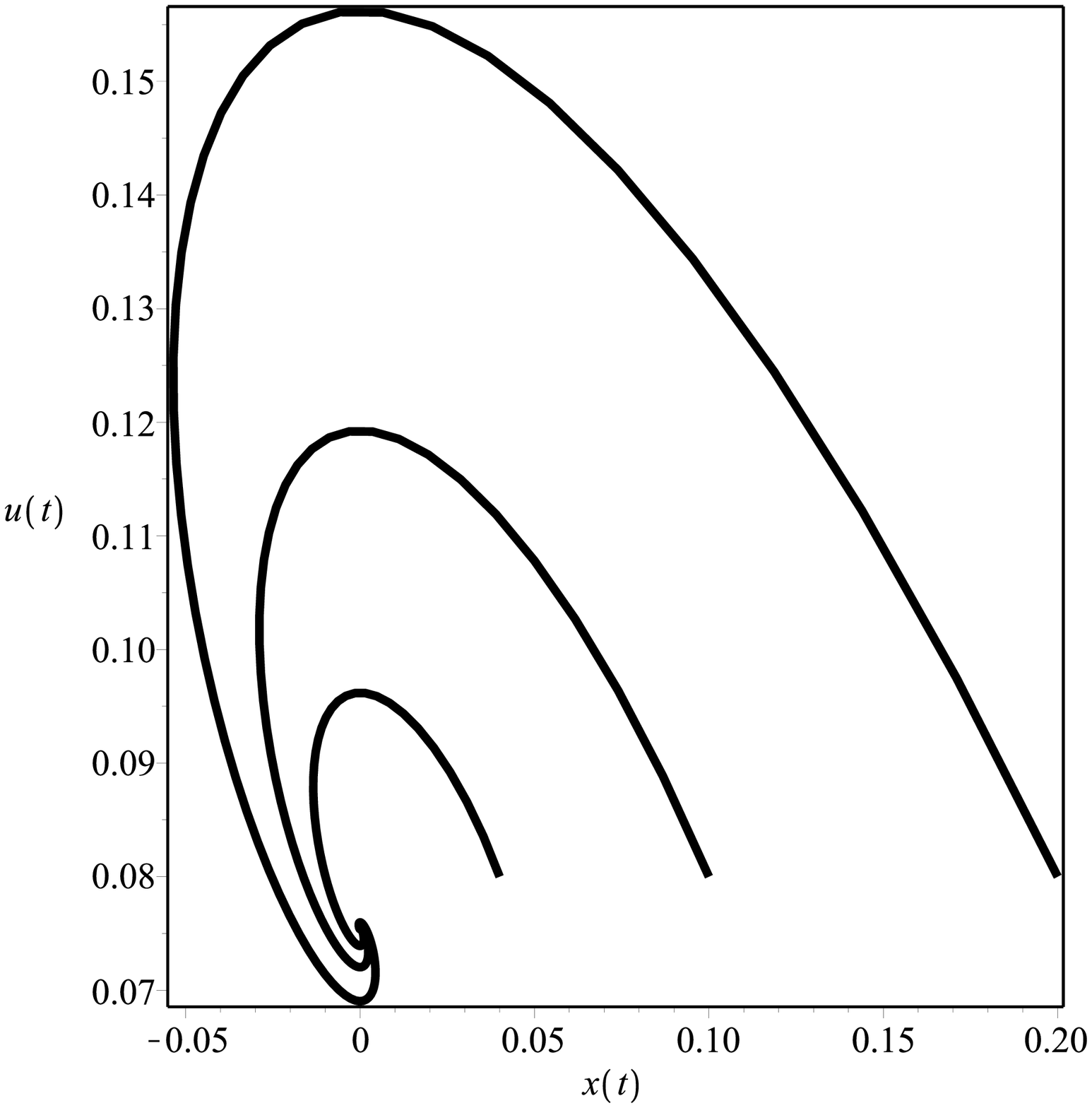}\vspace{2.5cm}\includegraphics{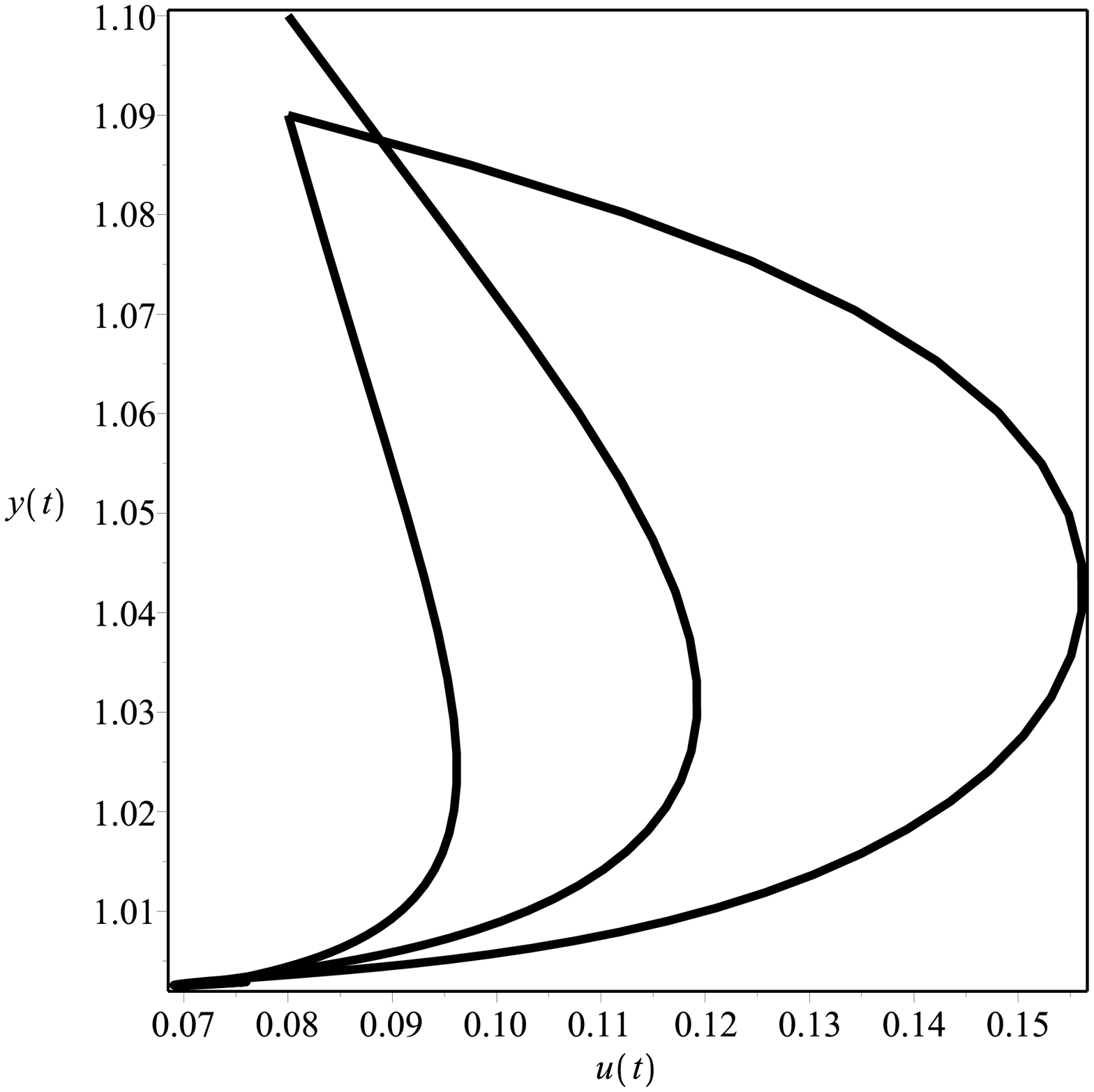} \caption{\small {From
left to right, the projections of the phase-space trajectories on
the $x-y$, $x-u$ and $u-y$ planes with $\xi=0.5$, $\lambda=0.6$ and
$\alpha=2$ for $Q=0$. For these values of the parameters, point
$A_{2}$ is a stable attractor of the model. }}

\end{center}
\end{figure}

\begin{figure}[htp]
\begin{center}
\includegraphics{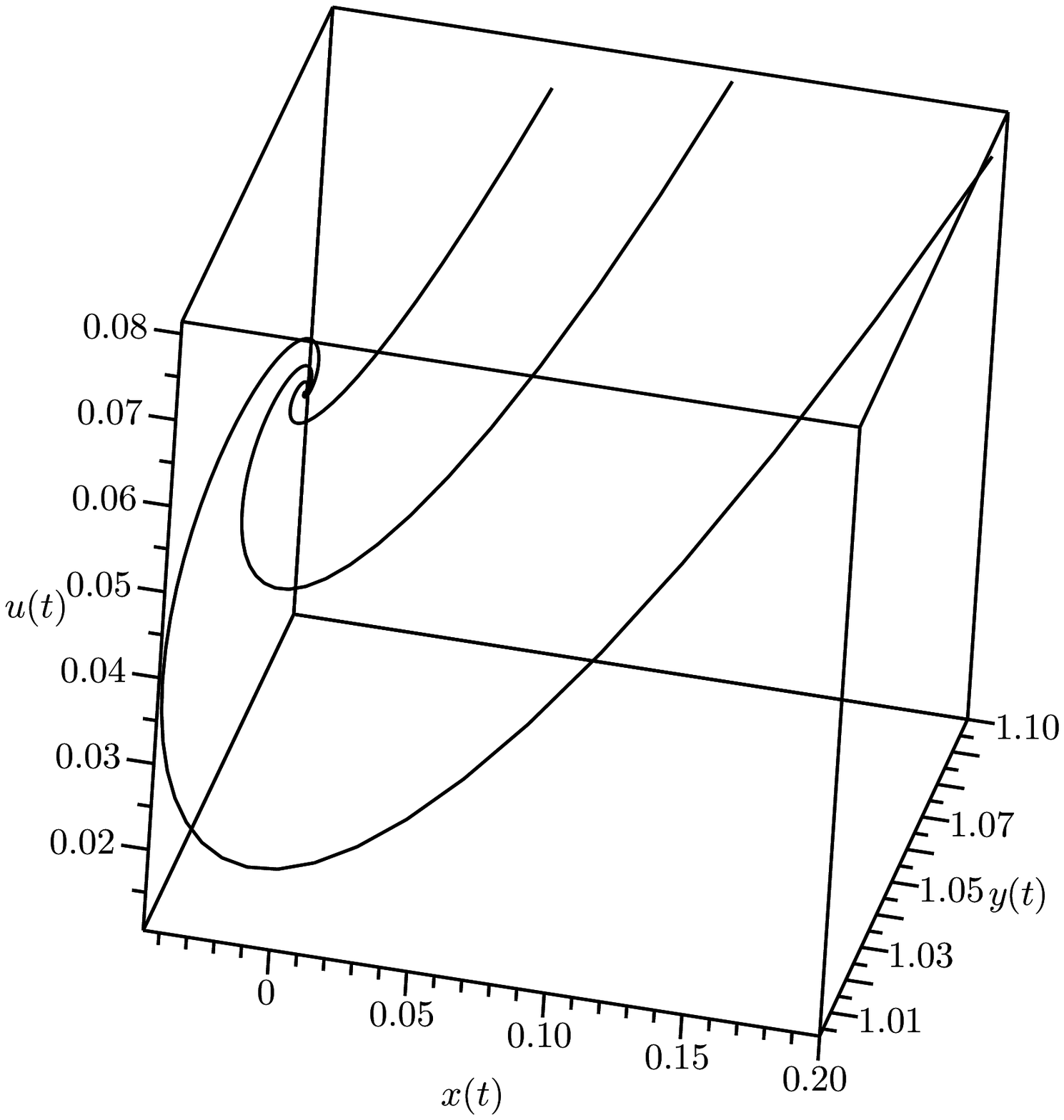} \vspace{2.5cm}\includegraphics{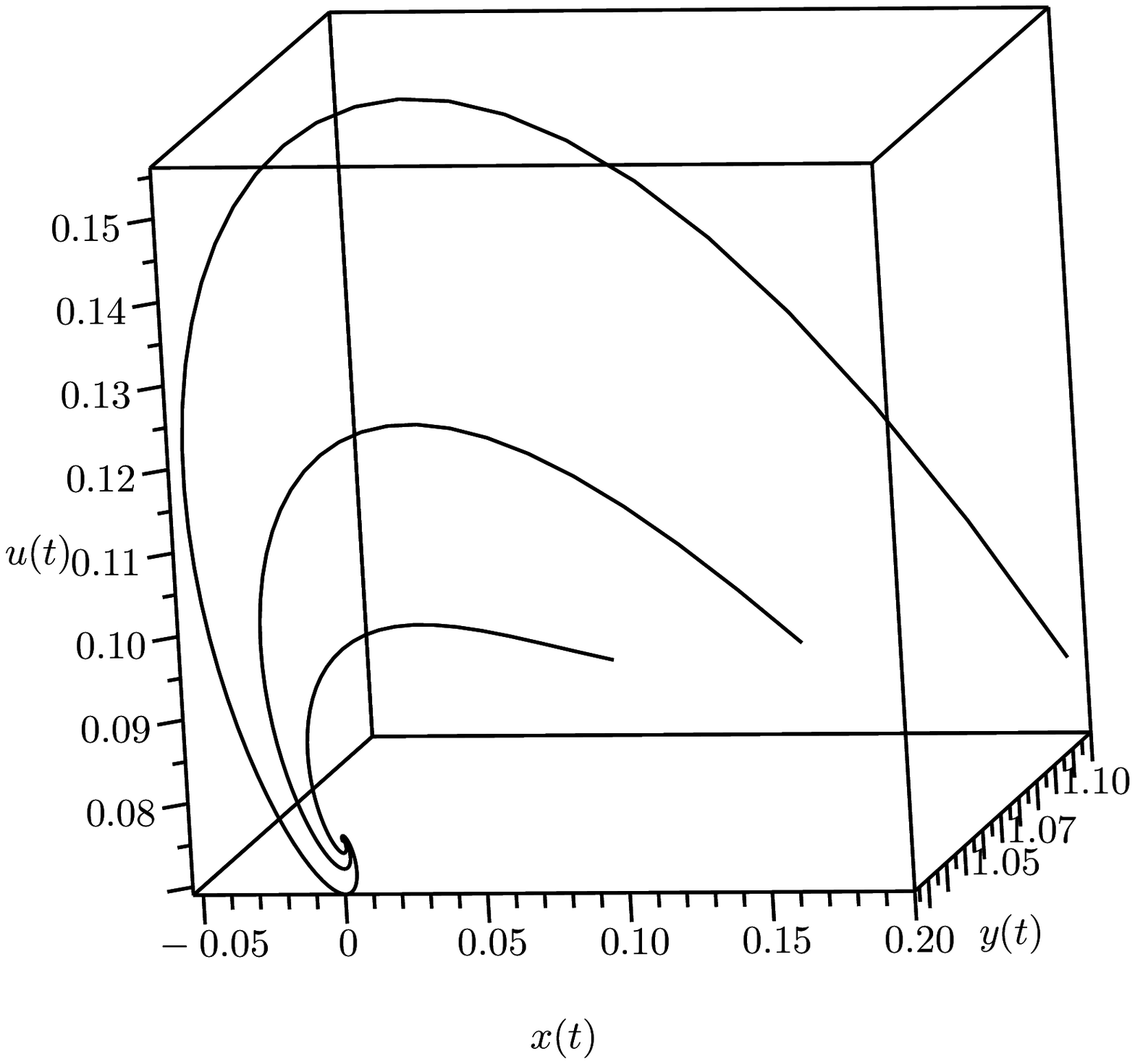}\vspace{2.5cm}

\caption{\small {3-dimensional phase-space trajectories of the model
for $Q=0$ with stable attractors $A_{1}$ (left) and $A_{2}$ (right)
. The values of the parameters are those mentioned in figure 1 and 2
respectively. }}
\end{center}
\end{figure}

\begin{figure}[htp]
\begin{center}
\includegraphics{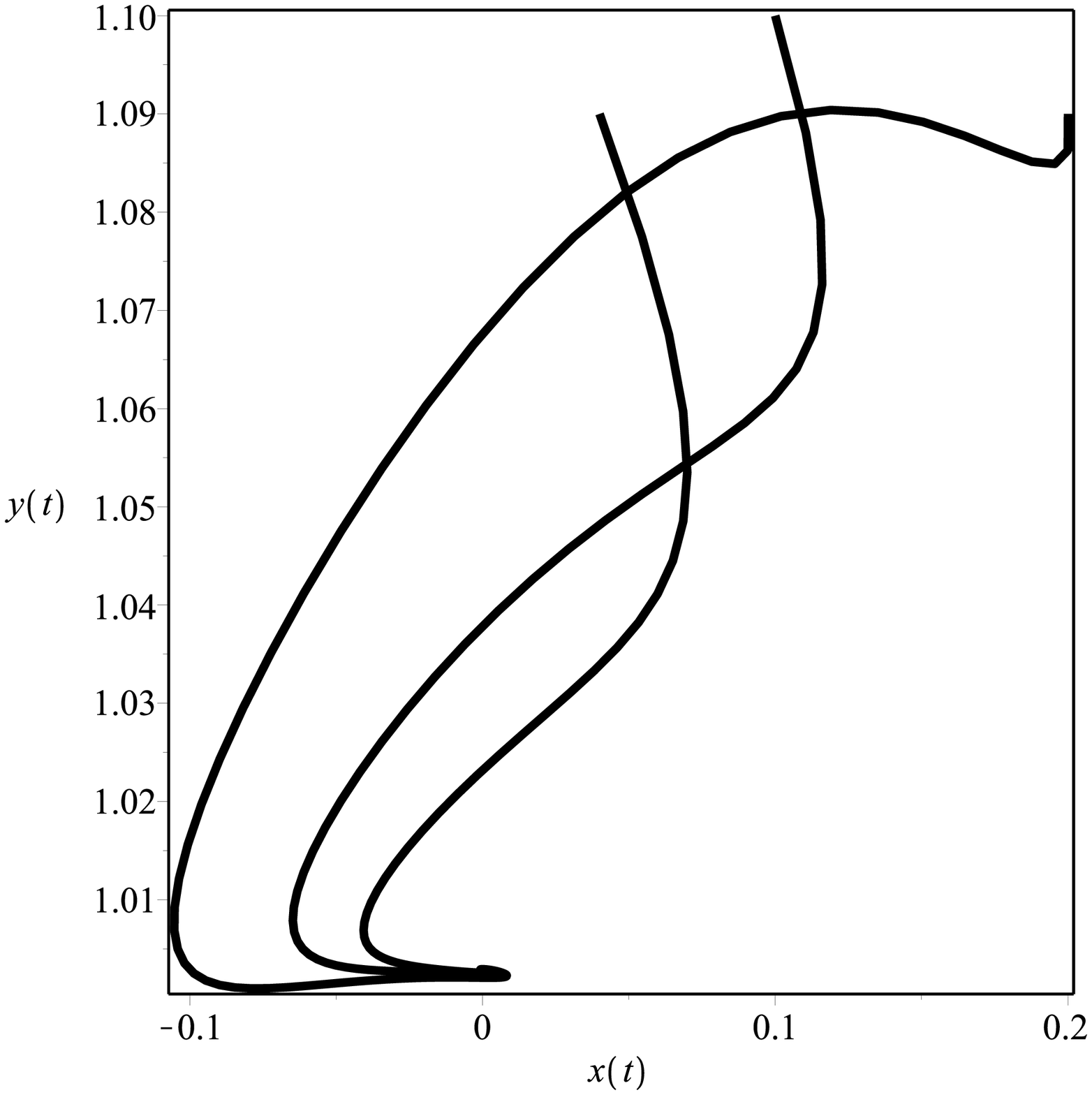} \vspace{1.5cm}\includegraphics{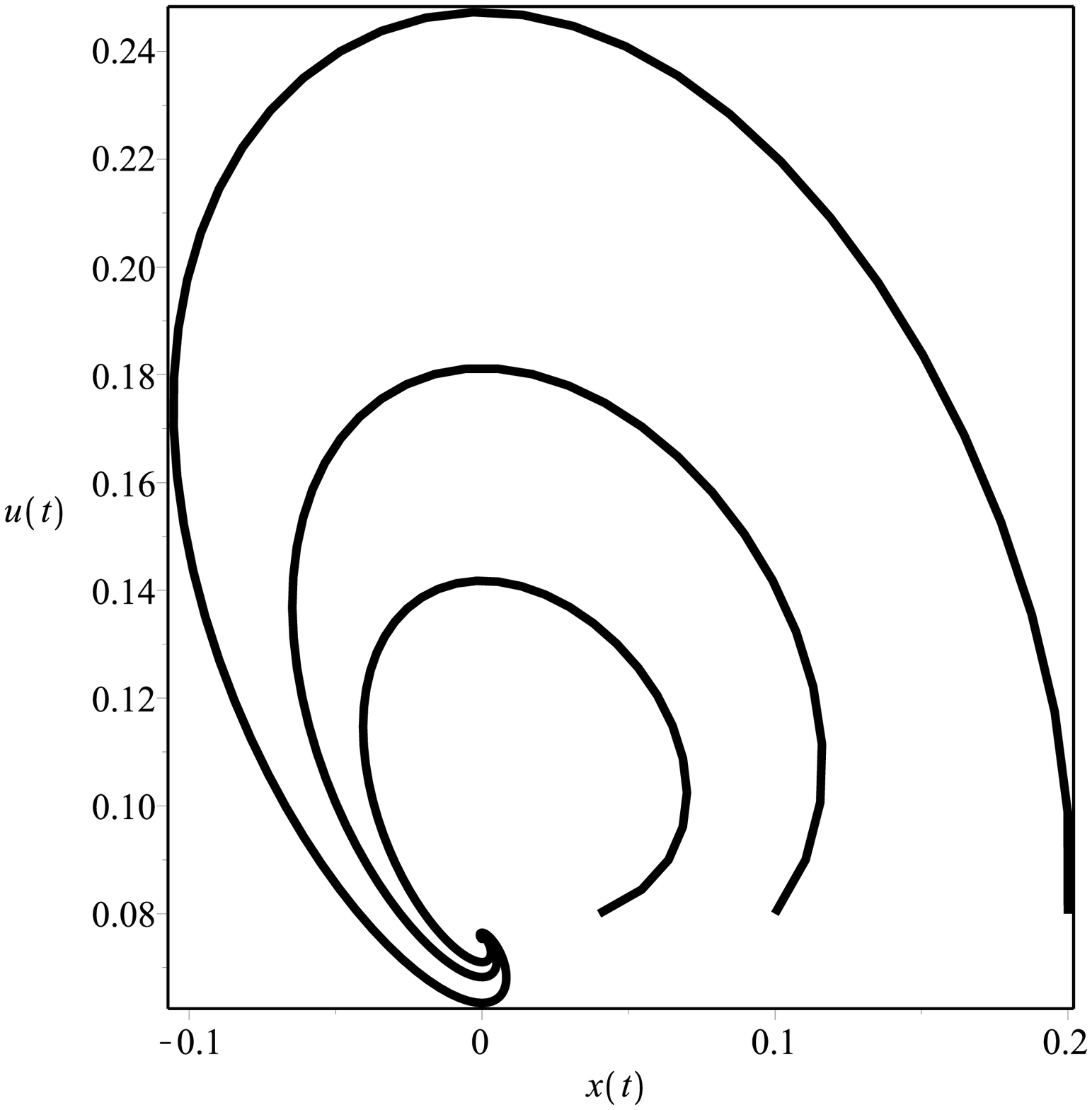}\vspace{1.5cm}\includegraphics{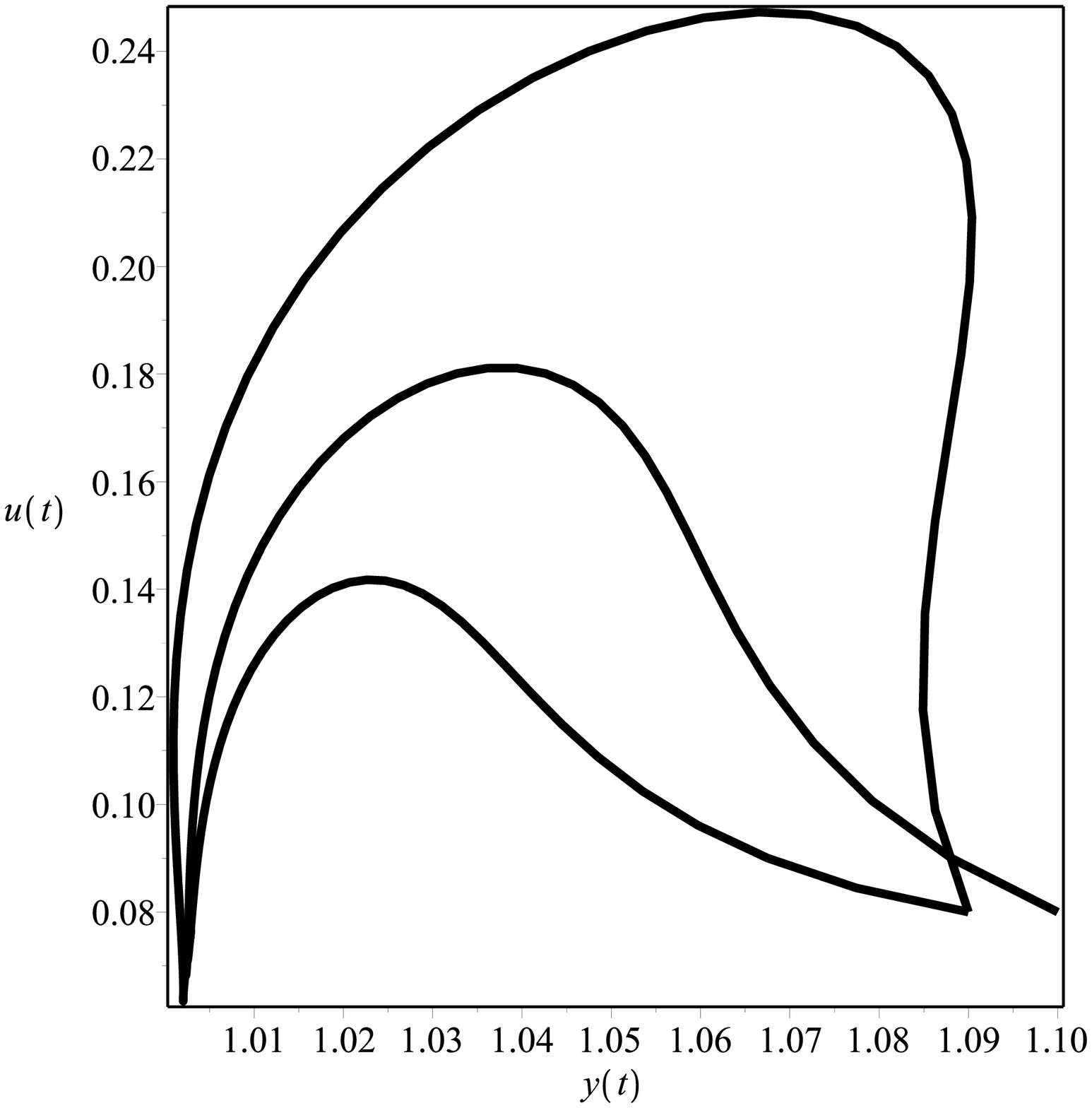} \vspace{4.5cm}

\caption{\small {From left to right, the projections of the
phase-space trajectories on the $x-y$, $x-u$ and $u-y$ planes with
$\xi=0.5$, $\lambda=0.6$, $\alpha=2$ and $\beta=1.5$ for
$Q=\beta\kappa\rho_{m}\dot{\phi}$. For these values of the
parameters, point $B_{2}$ is a stable attractor of the model. }}
\end{center}
\end{figure}

\begin{figure}[htp]
\begin{center}
\includegraphics{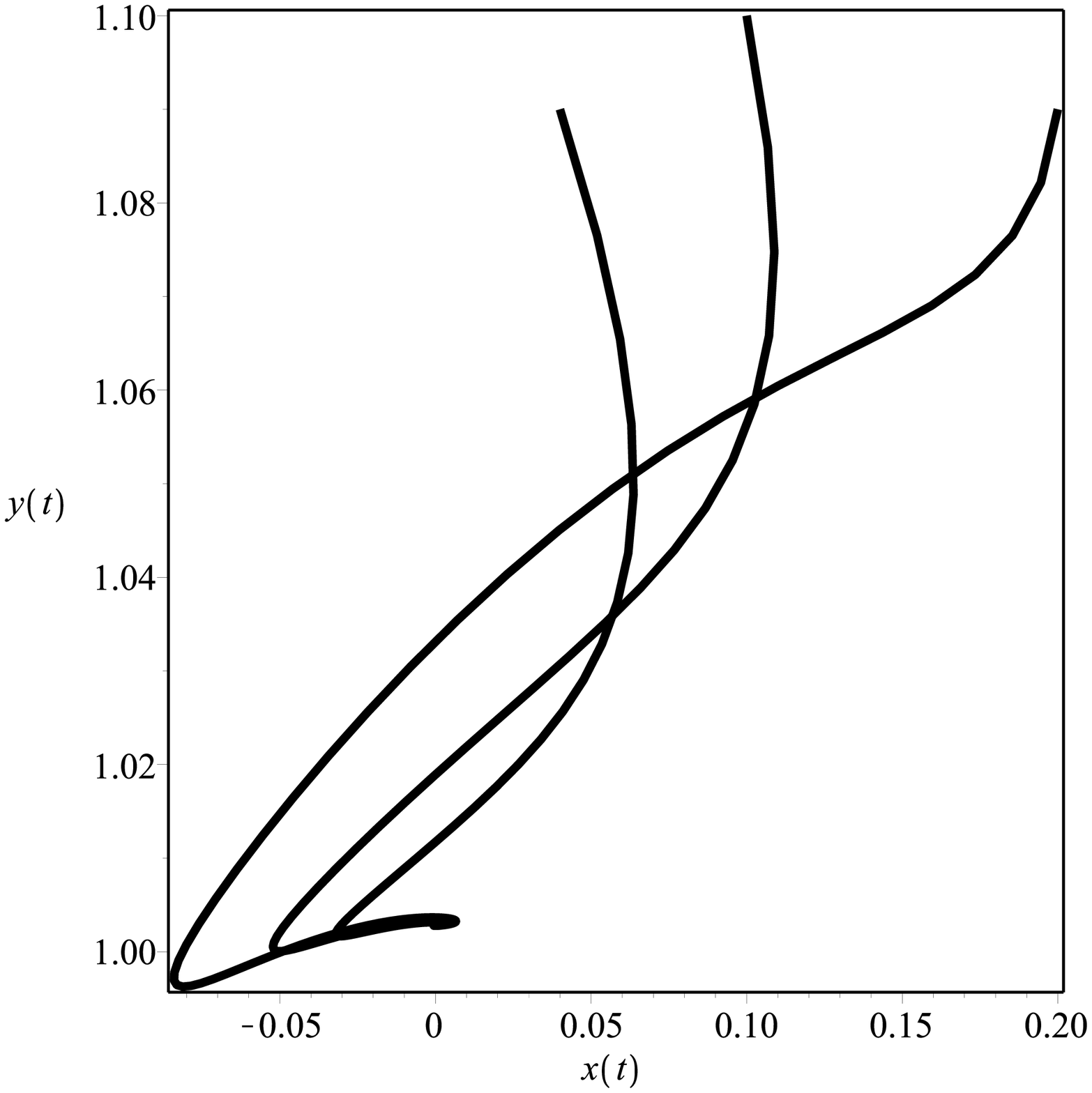} \vspace{1.5cm}\includegraphics{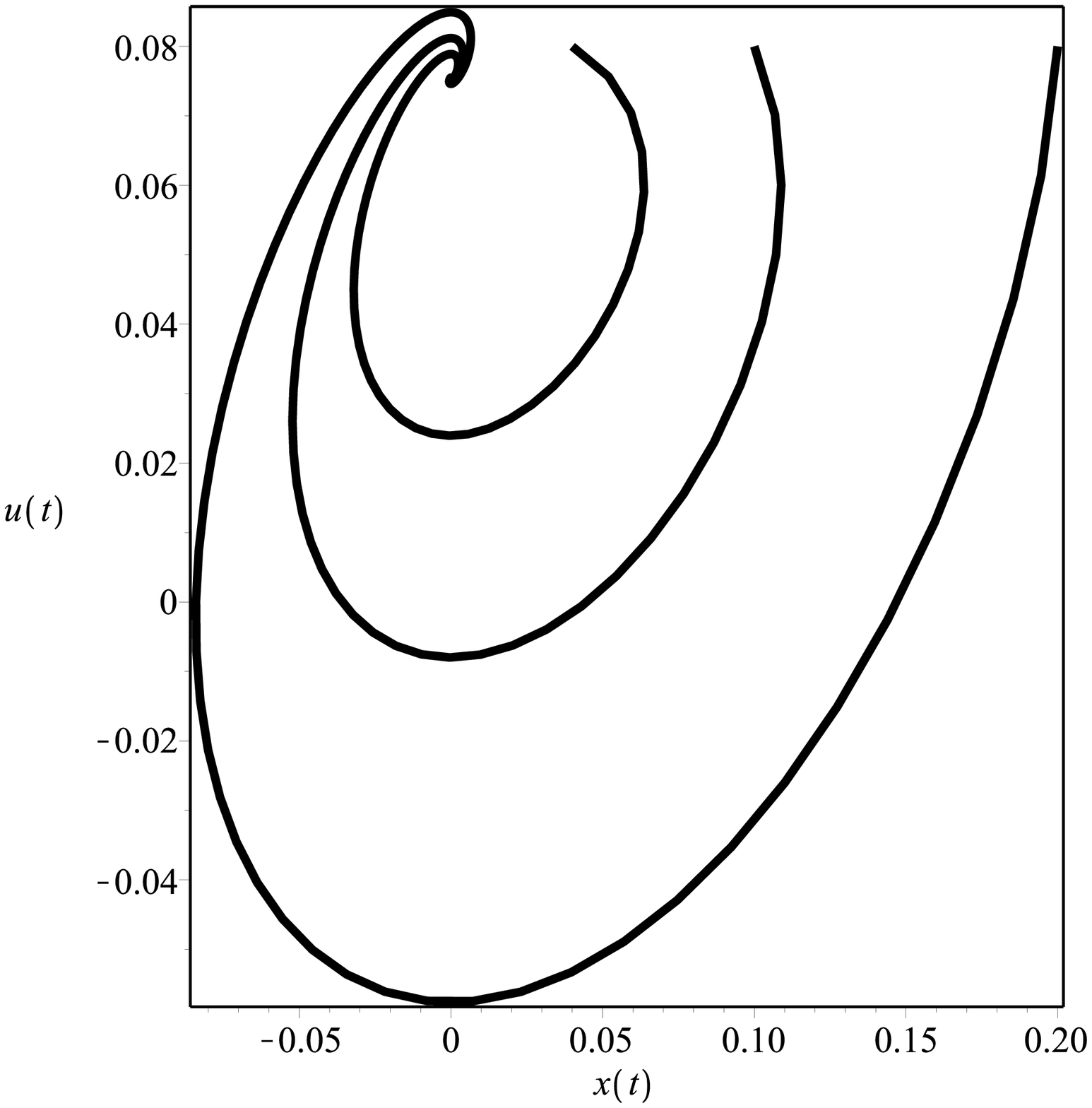}\vspace{1.5cm}\includegraphics{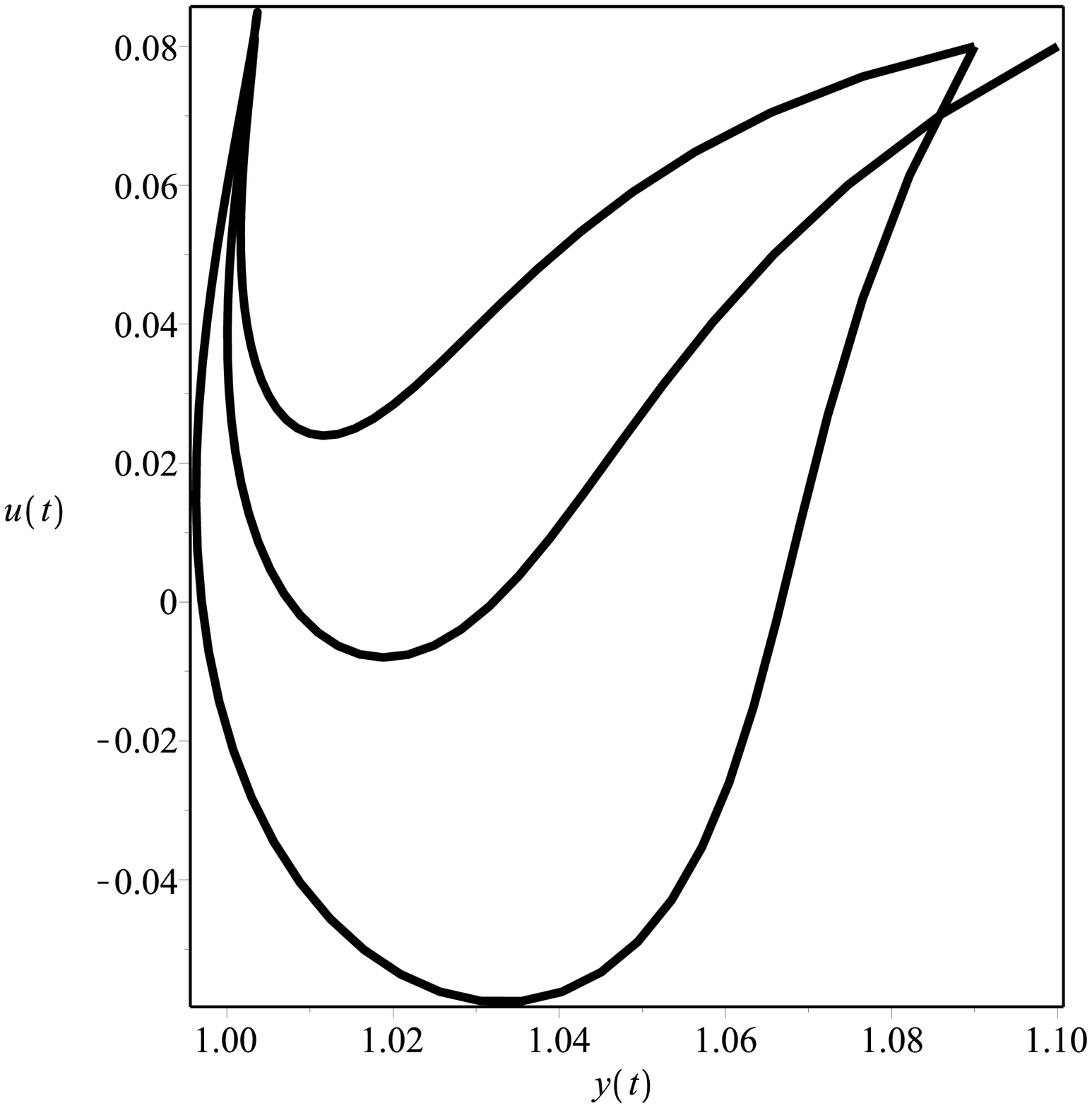} \vspace{4.5cm}

\caption{\small {From left to right, the projections of the phase
space trajectories on the $x-y$, $x-u$ and $u-y$ planes with
$\xi=0.5$, $\lambda=-0.6$, $\alpha=-2$ and $\beta=1.5$ for
$Q=\beta\kappa\rho_{m}\dot{\phi}$. For these values of the
parameters, point $B_{3}$ is a stable attractor of the model. }}
\end{center}
\end{figure}

\begin{figure}[htp]
\begin{center}
\includegraphics{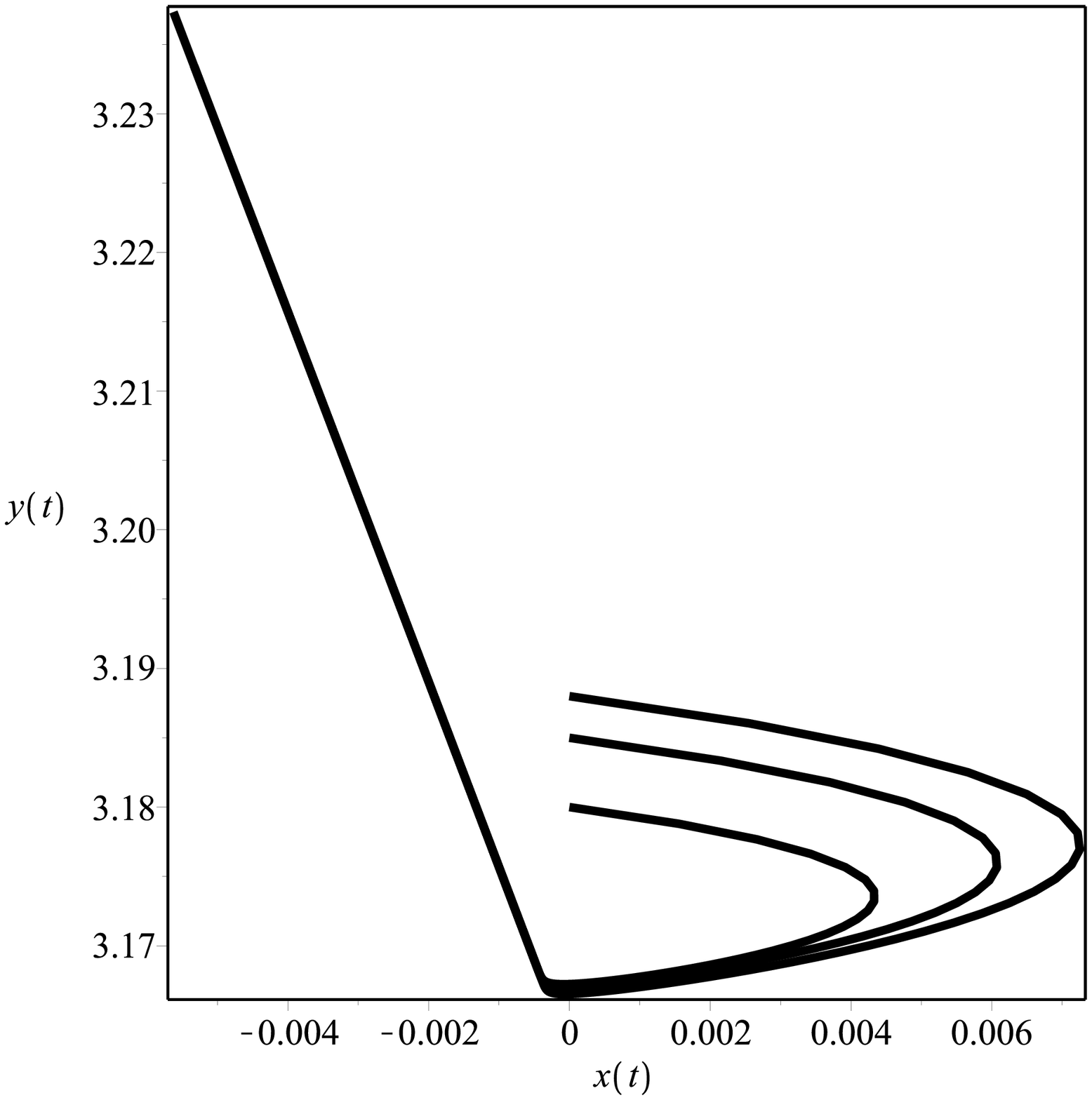} \vspace{1.5cm}\includegraphics{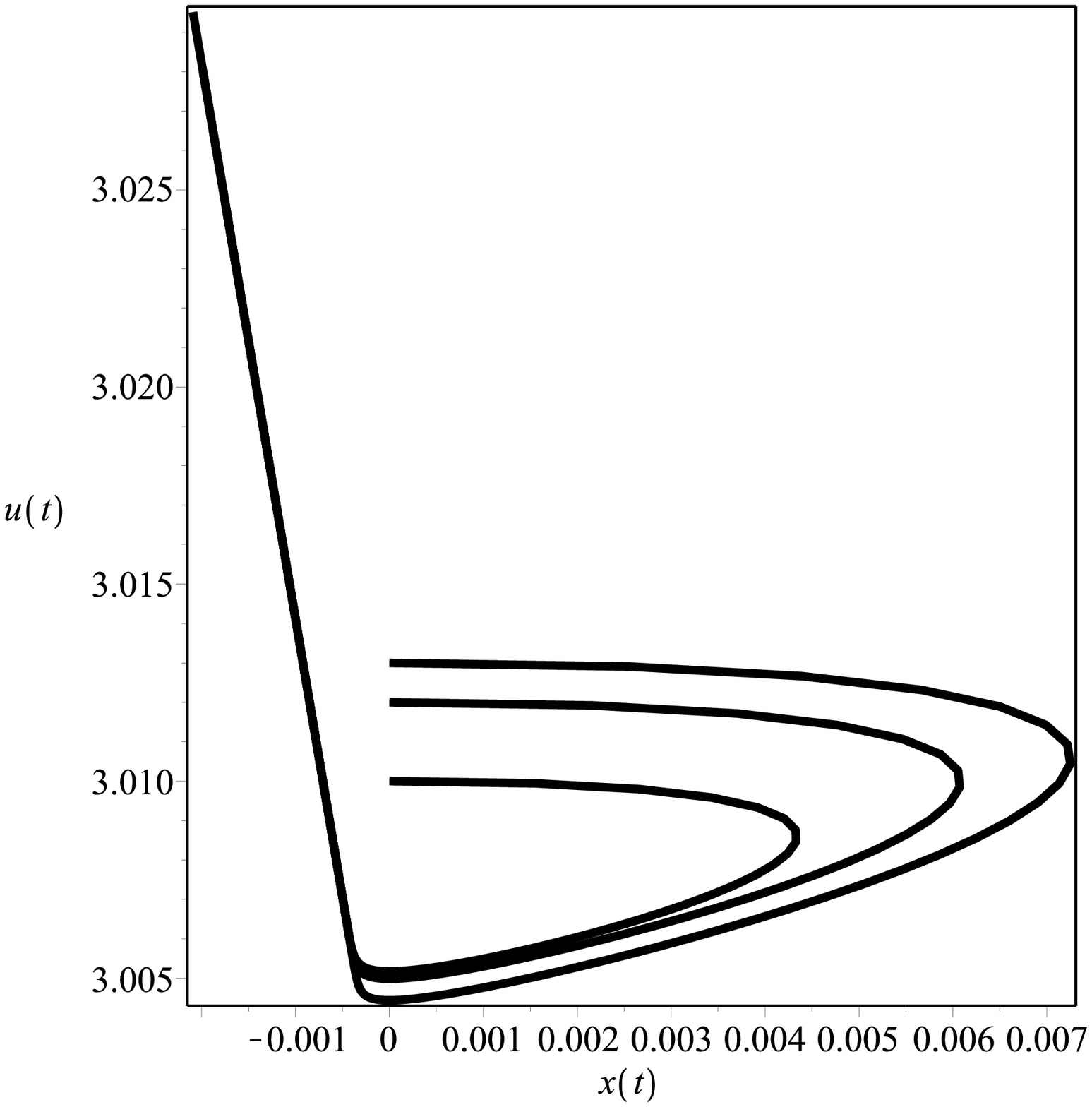}\vspace{1.5cm}\includegraphics{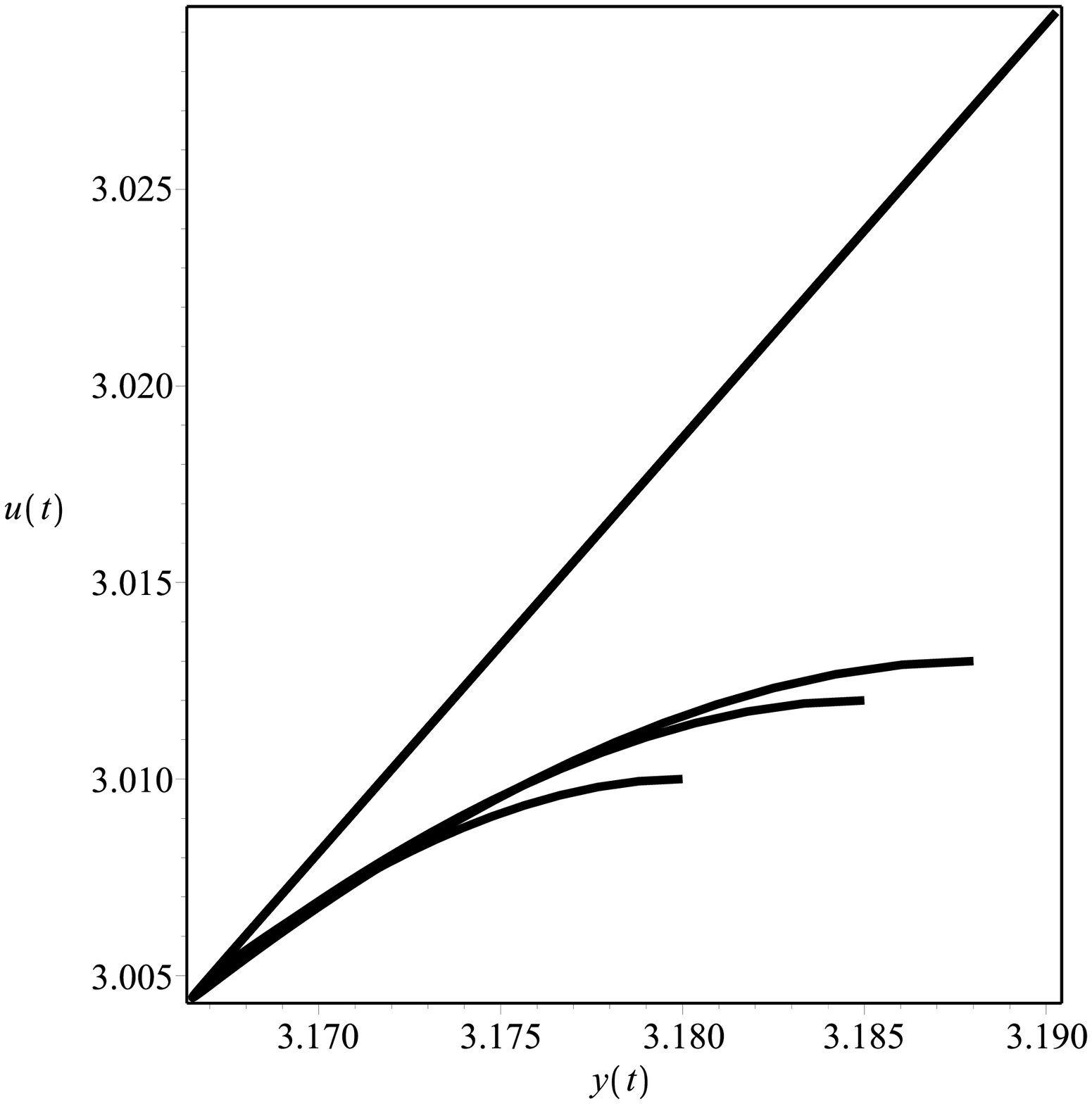} \vspace{4.5cm}

\caption{\small {From left to right, the projections of the phase
space trajectories on the $x-y$, $x-u$ and $u-y$ planes with
$\xi=0.5$, $\lambda=-0.6$, $\alpha=-2$ and $\beta=1.5$ for
$Q=\beta\kappa\rho_{m}\dot{\phi}$. For these values of the
parameters, point $B_{4}$ is a stable attractor of the model. }}
\end{center}
\end{figure}

\begin{figure}[htp]
\begin{center}
\includegraphics{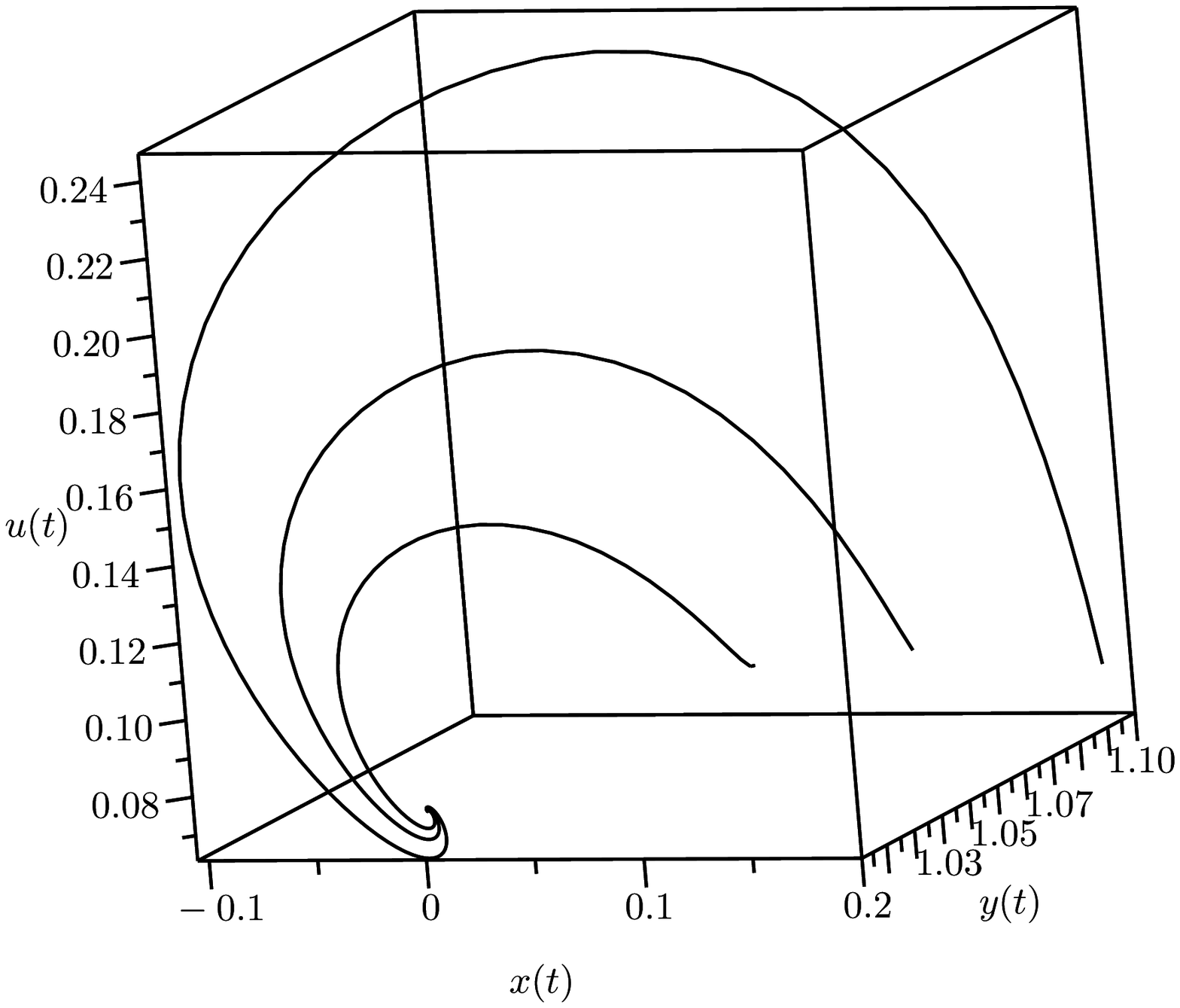} \vspace{1.5cm}\includegraphics{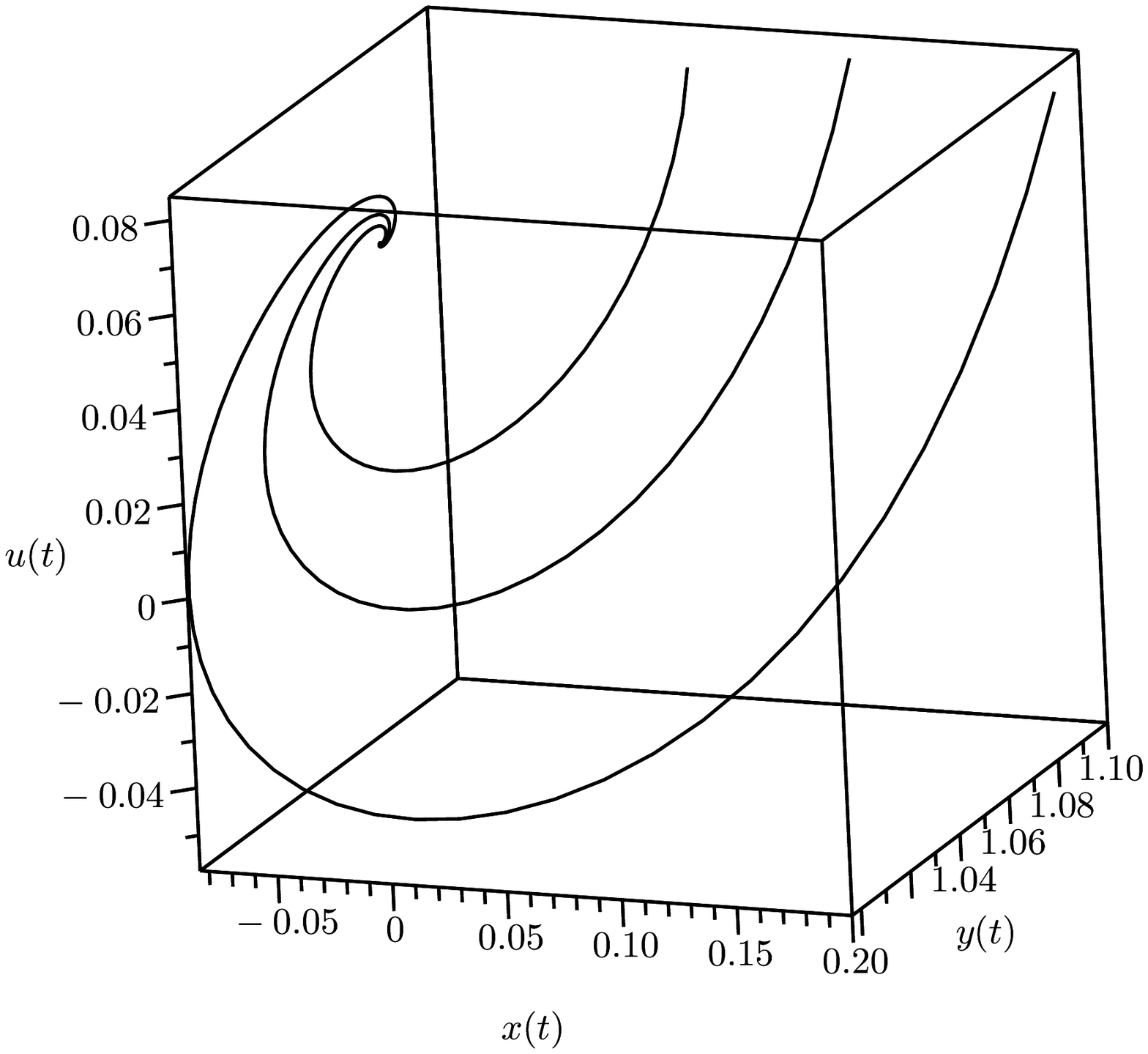}\vspace{1.5cm}\includegraphics{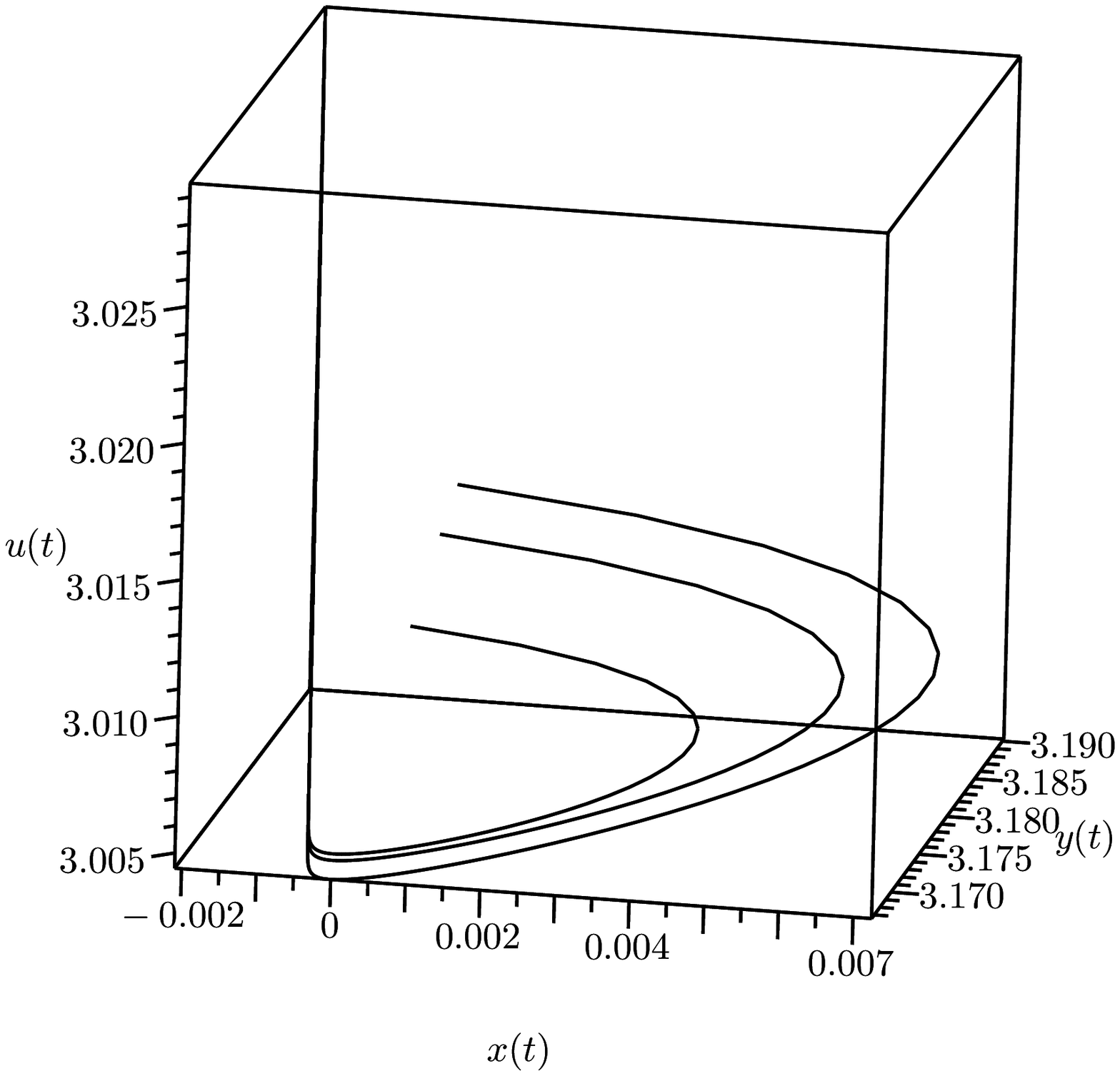} \vspace{4.5cm}

\caption{\small {3-dimensional phase-space trajectories of the model
for $Q=\beta\kappa\rho_{m}\dot{\phi}$ with stable attractors $B_{2}$
(left), $B_{3}$ (middle) and $B_{4}$ (right). The values of the
parameters are those mentioned in figure 4, 5 and 6 respectively.}}
\end{center}
\end{figure}

\newpage
\section{Appendix: Perturbation Matrix Elements}
The elements of $3 \times 3$ matrix $M$ of the linearized
perturbation equations for the real and physically meaningful
critical points $(x_{c}, y_{c}, u_{c})$ of the autonomous system
(20)- (22) read,
\begin{equation}
M_{11}=3\nu_{c}^{2}\Big(\frac{\sqrt{3}}{2}\lambda
x_{c}y_{c}\big(2x_{c}^{2}+\nu_{c}^{2}(1+3x_{c}^{4})\big)-6\mu_{c}^{-2}\nu_{c}^{2}x_{c}^{2}-4\sqrt{3}\alpha\xi
u_{c}x_{c}\nu_{c}^{2}y_{c}^{-1}\Big)+\sqrt{3}\lambda x_{c}
y_{c}-3+\mathcal{M}_{11},
\end{equation}

\begin{equation}
M_{12}=\frac{\sqrt{3}}{4}\Big(\lambda\big(2x_{c}^{2}+\nu_{c}^{2}(1+3x_{c}^{4})\big)+8\alpha\xi
u_{c}\nu_{c}^{2}y_{c}^{-2}\Big)+\mathcal{M}_{12},
\end{equation}

\begin{equation}
M_{13}=-2\sqrt{3}\alpha\xi \nu_{c}^{2}y_{c}^{-1}+\mathcal{M}_{13},
\end{equation}

\begin{equation}
M_{21}=\frac{2y_{c}^{2}\big(\sqrt{3}\alpha\xi u_{c}+3 x_{c} y_{c}
\nu_{c}^{-2}\big)}{(2\xi u_{c}^{2}+1)}-\frac{\sqrt{3}\lambda
y_{c}^{2}}{2},
\end{equation}

\begin{equation}
M_{22}=\frac{2y_{c}\big(-\frac{9}{4}\mu_{c}^{-4}y_{c}+5\sqrt{3}\alpha\xi
x_{c}u_{c}\big)}{(2\xi u_{c}^{2}+1)}-\sqrt{3}\lambda
x_{c}y_{c}+\frac{3}{2},
\end{equation}

\begin{equation}
M_{23}=\frac{6\xi u_{c} y_{c}^{2}\big(-\frac{10\sqrt{3}}{3}\alpha\xi
u_{c}x_{c}+\mu_{c}^{-4}y_{c}^{2}\big)}{(2\xi
u_{c}^{2}+1)^{2}}+\frac{5\sqrt{3}\alpha\xi x_{c} y_{c}^{2}}{2\xi
u_{c}^{2}+1},
\end{equation}

\begin{equation}
M_{31}=\frac{\sqrt{3}\alpha
y_{c}}{2},\,\,\,\,M_{32}=\frac{\sqrt{3}\alpha
x_{c}}{2},\,\,\,\,M_{33}=0,
\end{equation}
where in the case of $Q=0$ we have
$\mathcal{M}_{11}=\mathcal{M}_{12}=\mathcal{M}_{13}=0$ and in the
case of $Q=\beta\kappa\rho_{m}\dot{\phi}$ we have

$$\mathcal{M}_{11}=4\sqrt{3}\beta\nu_{c}^{-2} x_{c} y_{c},$$
$$\mathcal{M}_{12}=2\sqrt{3}\beta\mu_{c}^{-2}(1+3x_{c}^{2}),$$
\begin{equation}
\mathcal{M}_{13}=-4\sqrt{3}\beta\xi u_{c} y_{c}^{-1}.
\end{equation}

 Examining the eigenvalues of the matrix $M$ for each critical
point, one determines its stability conditions. \\

\end{document}